\documentclass[twocolumn]{aastex62}
\graphicspath{{./}{figures/}}

\accepted{April 30, 2020}

\shorttitle{The SAMI-GS: stellar population gradients of central galaxies}
\shortauthors{Santucci et al.}

\begin{document}
	
	\title{The SAMI Galaxy Survey: stellar population gradients of central galaxies}

	\correspondingauthor{Giulia Santucci}
	\email{g.santucci@student.unsw.edu.au}
	
	\author[0000-0003-3283-4686]{Giulia Santucci}
	\affil{School of Physics, University of New South Wales, NSW 2052, Australia}
	
	\author{Sarah Brough}
	\affil{School of Physics, University of New South Wales, NSW 2052, Australia}
	\affil{Australia ARC Centre of Excellence for All Sky Astrophysics in 3 Dimensions (ASTRO 3D)}

	\author{Nicholas Scott}
	\affil{Sydney Institute for Astronomy, School of Physics, University of Sydney, NSW 2006, Australia}
	\affil{Australia ARC Centre of Excellence for All Sky Astrophysics in 3 Dimensions (ASTRO 3D)}	
	
	\author{Mireia Montes}
	\affil{School of Physics, University of New South Wales, NSW 2052, Australia}
	
	\author{Matt S. Owers}
	\affil{Department of Physics and Astronomy, Macquarie University, NSW 2109, Australia}
	\affil{Astronomy, Astrophysics and Astrophotonics Research Centre, Macquarie University, Sydney, NSW 2109, Australia}
	
	\author{Jesse van de Sande}
	\affil{Sydney Institute for Astronomy, School of Physics, University of Sydney, NSW 2006, Australia}
	\affil{Australia ARC Centre of Excellence for All Sky Astrophysics in 3 Dimensions (ASTRO 3D)}
	
	\author{Joss Bland-Hawthorn}
	\affil{Sydney Institute for Astronomy, School of Physics, University of Sydney, NSW 2006, Australia}
	
	\author{Julia J. Bryant}
	\affil{Sydney Institute for Astronomy, School of Physics, University of Sydney, NSW 2006, Australia}
	\affil{Australia ARC Centre of Excellence for All Sky Astrophysics in 3 Dimensions (ASTRO 3D)}
	\affil{Australian Astronomical Optics, AAO-USydney, School of Physics, University of Sydney, NSW 2006, Australia}
	
	\author{Scott  M. Croom}
	\affil{Sydney Institute for Astronomy, School of Physics, University of Sydney, NSW 2006, Australia}
	\affil{Australia ARC Centre of Excellence for All Sky Astrophysics in 3 Dimensions (ASTRO 3D)}

	\author{Ignacio Ferreras}
	\affil{Department of Physics and Astronomy, University College London, Gower St, London WC1E 6BT, UK}
	\affil{Instituto de Astrof\'isica de Canarias, Calle V\'a L\'actea s/n, E-38205, La Laguna, Tenerife, Spain}
	\affil{Departamento de Astrofísica, Universidad de La Laguna (ULL), E-38206 La Laguna, Tenerife, Spain}
	
	\author{Jon S. Lawrence}
	\affil{Australian Astronomical Optics - Macquarie, Macquarie University, NSW 2109, Australia}

	\author{\'Angel R. L\'opez-S\'anchez}
	\affil{Australian Astronomical Optics - Macquarie, Macquarie University, NSW 2109, Australia}
    \affil{Department of Physics and Astronomy, Macquarie University, NSW 2109, Australia}
	\affil{Australia ARC Centre of Excellence for All Sky Astrophysics in 3 Dimensions (ASTRO 3D)}

	\author{Samuel N. Richards}
	\affil{SOFIA, USRA, NASA Ames Research Center, Building N232, M/S 232-12, P.O. Box 1, Moffett Field, CA 94035-0001, USA}
	
	
	
	\begin{abstract}
		We examine the stellar population radial gradients (age, metallicity and [$\alpha/$Fe]) of 96 passive central 
		galaxies up to $\sim 2 R_e$ in the SAMI Galaxy Survey. The targeted groups have a halo mass range spanning from $11 < \log(M_{200}/M_{\odot}) < 15$. The main goal of this work is to determine whether central galaxies have different stellar population properties when compared to similarly massive satellite galaxies. For the whole sample we find negative metallicity radial gradients, which show evidence of becoming shallower with increasing stellar mass. The age and [$\alpha$/Fe] gradients are slightly positive and consistent with zero, respectively. The [$\alpha$/Fe] gradients become more negative with increasing mass, while the age gradients do not show any significant trend with mass. We do not observe a significant difference between
		the stellar population gradients of central and satellite galaxies, at fixed stellar mass. The mean metallicity gradients are $\overline{\Delta [Z/H]/\Delta \log(r/R_e)} = -0.25 \pm 0.03$ for central galaxies and $\overline{\Delta [Z/H]/\Delta \log(r/R_e)} = -0.30 \pm 0.01$ for satellites. The mean age and [$\alpha$/Fe] gradients are consistent
		between central and satellite galaxies, within the uncertainties, with a mean value of $\overline{\Delta \textrm{log (Age/Gyr)}/\Delta \log(r/R_e)} = 0.13 \pm 0.03$ for centrals and $\overline{\Delta \textrm{log (Age/Gyr)}/\Delta \log(r/R_e)} = 0.17 \pm 0.01$ for satellites and $\overline{\Delta [\alpha/Fe]/\Delta \log(r/R_e)} = 0.01 \pm 0.03$ for centrals and $\overline{\Delta [\alpha/Fe]/\Delta \log(r/R_e)} = 0.08 \pm 0.01$ for satellites. The stellar population gradients of central and satellite galaxies show no difference as a  function  of  halo  mass.
		This evidence suggests that the inner regions of central passive galaxies form in a similar fashion to those of satellite passive galaxies, in agreement with a two-phase formation scenario.
	\end{abstract}
	
	
	
\section{Introduction} \label{sec:intro}
In the hierarchical galaxy formation paradigm, galaxy assembly and the growth of dark matter halos are closely linked. The central galaxies in dark matter halos are generally the brightest galaxies in those systems (also known as Brightest Cluster Galaxies and Brightest Group Galaxies: BCGs and BGGs, here referred to as central galaxies). Due to their privileged position at the bottom of the gravitational potential, they are predicted to have undergone a higher rate of mergers and, therefore, to generally be more massive than other galaxies in those systems \citep[e.g.,][]{DeLucia2007}.
Despite being very luminous and relatively easy to detect, the formation history of central galaxies is complex and not yet fully understood \citep[e.g.,][]{Lin2013, Lidman2013, Oliva-Altamirano2014, Davies2019}.

Luckily, the stellar populations of galaxies give an insight into their formation histories. The fraction of stars with a given age is a record of the star formation history of the galaxy, while the metallicity gradient is strongly correlated with the galaxy's assembly history and the abundance of alpha-elements relative to iron elements ([$\alpha$/Fe]) gives information on the star formation timescale in the different regions of the galaxy.

In particular, studying how the stellar populations vary with radius can shed light on whether the evolutionary histories of central galaxies are different from those of satellite galaxies, which are not in a privileged position in the gravitational well.

Massive galaxies at $z \sim 2$ are already found to be quiescent, but compact, with
an effective radius ($R_e$) half of the size of galaxies of similar mass in the local universe \citep[e.g.,][]{Daddi2005, Trujillo2006, vanDokkum2008}. Hydrodynamical-zoom cosmological simulations found that massive galaxies likely form in a two-phase formation scenario \citep[e.g.,][]{Naab2009}. During the first phase, at high redshift, they grow by a rapid episode of in-situ star formation, resulting in compact massive systems. The initial stellar metallicity gradient is set by the initial episode of star formation, with the metallicity decreasing outward in the galaxy \citep{Larson1974, Thomas2005}. From this process, we would expect steep, negative metallicity gradients and flat age radial profiles. 
After $z \approx 2$, these massive compact galaxies ($\log_{10} (M_*/ M_{\odot}) > 10.5$) are predicted to be quiescent and grow mostly by accreting mass through galaxy interactions that add stars to their outskirts \citep{Oser2010, Naab2009,Hopkins2009}.

In the simulations of \cite{Hirschmann2015} and \cite{Cook2016}, galaxies that have experienced a greater number of interactions show flat or slightly positive age gradients, due to old stars being added in the outer regions, and shallower metallicity gradients, produced by the mixing of the different metallicities of the accreted stars. In these simulations, the contributions from accretion start to be visible beyond 2 $R_e$ (seen observationally by \citealt{Coccato2010, Montes2014}). By observing the radial stellar population gradients of galaxies we can infer their assembly history.

Several observational studies have explored the stellar population radial profiles of central galaxies, using long-slit spectroscopy: \cite{Brough2007} and \cite{Loubser2012} found that the age and [$\alpha$/Fe] gradients are mostly shallow and similar to those of other ETGs in high-density environments. They also found a wide range of metallicity gradients, suggesting that these galaxies had very different assembly histories.

Recently, substantial progress on our understanding of the stellar populations of early-type galaxies (ETGs) has been made thanks to the use of large Integral Field Spectroscopy (IFS) surveys such as SAURON (Spectroscopic Areal Unit
for Research on Optical Nebulae; \citealt{deZeeuw2002}), ATLAS\(^{3D}\) \citep{Cappellari2011}, CALIFA (Calar Alto Legacy Integral Field Array survey; \citealt{Sanchez2012}), MASSIVE \citep{Ma2014}, MaNGA (Mapping Nearby Galaxies at Apache Point Observatory; \citealt{Bundy2015}) and the SAMI (Sydney-Australian-Astronomical-Observatory Multi-object Integral-Field Spectrograph) Galaxy Survey \citep{Croom2012, Bryant2015}. IFS enables the mapping of stellar populations across individual galaxies, rather than obtaining gradients from only one axis of the galaxy as long-slit observations do.

Despite the increasing number of surveys, the studies of the stellar population gradients of ETGs are still not in agreement. For example, \cite{Kuntschner2010} found, studying 48 ETGs in the SAURON survey, evidence for gradually shallower metallicity gradients ($\Delta$[Z/H]/$\Delta \log$ (r/$R_e$) $>$ $-$0.3, measured within 1 $R_e$) with increasing stellar mass, for galaxies with masses greater than $10^{10.3}  M_{\odot}$; and flat age gradients for the general population of ETGs. Similar results for metallicity gradients (within 1 $R_e$) were found by \cite{Li2018}, in MaNGA galaxies with masses ranging from $10^9$ and $10^{12.3}  M_{\odot}$. However, \cite{Goddard2017a} studying MaNGA early-type galaxies with stellar masses ranging from $10^9$ and $10^{11.5}  M_{\odot}$, concluded that light-weighted metallicity gradients (within 1.5 $R_e$) become slightly more negative with increasing stellar mass (with the relation having a slope of $-0.04 \pm 0.05$). \cite{Zheng2017} found weak or no correlation between metallicity gradients and stellar mass with the same survey ($\approx$ 570 MaNGA ETGs with stellar masses ranging from $10^{8.5}$ and $10^{11.5}  M_{\odot}$; consistent with \citealt{Goddard2017a}). The differences, however, can be explained in most cases by differences in sample selection (how the ETGs were selected and the galaxy stellar mass range studied). In the particular case of \cite{Goddard2017a} and \cite{Li2018} the differences are also likely to be due to the different definition of gradient used - linear radial fits ($\Delta$[Z/H]/$\Delta$ (r/$R_e$); \citealt{Goddard2017a}) compared to logarithmic radial fits ($\Delta$[Z/H]/$\Delta \log$ (r/$R_e$); \citealt{Li2018}).

The MASSIVE survey found massive ETGs ($M_* > 4 \times 10^{11} \  M_\odot$) to have shallow metallicity gradients ($\Delta$ [Z/H]/$\Delta \log$ (r/$R_e$) = -0.3 $\pm$ 0.1) and no significant age or [$\alpha$/Fe] abundance ratio gradients \citep{Greene2015}. Moreover, extending the gradients to the outer regions of their galaxies (up to 3 $R_e$), they found  shallow negative metallicity gradients (median $\Delta$ [Z/H]/$\Delta \log$ (r/$R_e$) = -0.26)  and nearly flat [$\alpha$/Fe] gradients (median $\Delta$ [$\alpha$/Fe]/$\Delta \log$ (r/$R_e$) =-0.03; \citealt{Greene2019}).

Focussing on central galaxies, \cite{Oliva-Altamirano2015} found  shallow metallicity gradients ($\Delta$[Z/H]/$\Delta \log$ (r/$R_e$) $\gtrsim$ $-$0.3) and age gradients consistent with zero for 9 BCGs in the local universe, similar to other ETGs of the same mass.

There is also no agreement on the role of environment in stellar population gradients. Variation between low-density environments and high-density environments are expected since higher density regions are predicted to collapse earlier. This has been observed previously, by \citealt{LaBarbera2011} who found that ETGs, with stellar masses greater than $10^{10.5}  M_{\odot}$, in group environments have more positive age gradients and more negative metallicity gradients compared to field ETGs). However, \cite{Goddard2017a} found no trend of stellar population gradients  with environment with any of three different definitions of environment ($N^{th}$ nearest neighbour local number density, gravitational tidal strength, and
classifying between central and satellite galaxies). In contrast, \cite{Greene2015} found a marginal steepening of metallicity gradients (in gradients extending to 2.5 $R_e$) for satellite galaxies in cluster environments, compared to satellites in lower-mass halos. This is consistent with results for the SAMI Galaxy Survey from \cite{Ferreras2019}.

The results to date are still contradictory and do not provide clear understanding of the role that stellar mass or environment plays in shaping stellar population gradients. This is likely due to the fact that the samples studied to date have generally been relatively small and the full parameter space of galaxy stellar mass and environment has not been studied in a homogeneous, statistically-significant sample. In particular, studies focusing on central early-type galaxies have only constituted a few tens of galaxies.

In this paper we will investigate whether the evolutionary histories of central galaxies are different from those of satellite galaxies by studying their stellar population gradients (up to 2 $R_e$) as a function of stellar mass in group and cluster environments. New SAMI Galaxy Survey data allows us to study a statistically significant number of galaxies in a range of environments for the first time.

In Section 2 we describe the sample of galaxies and the data available for this analysis; Section 3 outlines the procedure followed to define a consistent and reliable set of data; Section 4 presents our results that are then discussed in Section 5. Our conclusions are given in Section 6.
SAMI adopts a $\Lambda CDM$ cosmology with $\Omega_m=0.3$, $\Omega_{\Lambda} = 0.7$, and $H_0 = 70$ km s$^{-1}$ Mpc$^{-1}$.

\section{Observations} \label{sec:style}
The Sydney-AAO Multi-object Integral field spectrograph (SAMI) Galaxy Survey is a large, optical 
Integral Field Spectroscopy (\citealt{Bryant2015}) survey of low-redshift 
($0.04 < z < 0.095$) galaxies covering a broad range in stellar mass,  $7 < \log_{10} (M_*/ M_{\odot}) < 12$, 
morphology and environment. The sample, with $\approx$ 3000 galaxies, is selected from the Galaxy and Mass Assembly Survey (GAMA survey; \citealt{Driver2011}) regions (group galaxies), as well as eight additional 
clusters to probe higher-density environments \citep{Owers2017}.

The SAMI instrument \citep{Croom2012}, on the 3.9m Anglo-Australian telescope, consists of 13 ``hexabundles" 
\citep{Bland-Hawthorn2011, Bryant2014}, across a 1 degree field of view. In the typical configuration, 
12 hexabundles are used to observe 12 science targets, with the 13th one allocated to a secondary
standard star used for calibration. Moreover, SAMI also has 26 individual sky fibers, to enable 
accurate sky subtraction for all observations without the need to observe separate blank sky frames.

\subsection{IFU Spectra}
SAMI data consist of three-dimensional data cubes: two spatial dimensions and a third spectral dimension.

The wavelength coverage is from 3750 to 5750 \AA\ in the blue arm, and from 6300 to 7400 \AA\ 
in the red arm, with a spectral resolution of R = 1812 (2.65 \AA\ full-width half maximum; FWHM) and R = 4263 (1.61 \AA ~FWHM), respectively \citep{vandeSande2017}, so that two data cubes are produced for each galaxy target. Only the blue arm is used to determine stellar population parameters, since there are only a few absorption features in the red wavelength range useful to this end.

The SAMI data reduction is fully described in \cite{Allen2015} and \cite{Sharp2015}, and we give a brief summary here:\\
A dedicated SAMI PYTHON package (that incorporates the 2DFDR package\footnote{\href{http://www.aao.gov.au/science/software/2dfdr}{http://www.aao.gov.au/science/software/2dfdr}}; \citealt{Croom2004, Sharp2010}) is used to perform bias subtraction, flat field normalization, remove cosmic rays and calibrate the absolute flux, to reduce the raw SAMI observations \citep{Allen2015}. Each galaxy field was observed in a set of approximately seven 30 minute exposures, that are aligned together by fitting the galaxy position within each hexabundle with a two-dimensional Gaussian and by fitting a simple empirical model describing the telescope offset and atmospheric
refraction to the centroids. The exposures are then combined to produce a spectral cube with regular $0.5^{\prime\prime}$ spaxels, with a median seeing of $2.1^{\prime\prime}$.
More details for Data Release 1 and Data Release 2 reduction can be found in \cite{Green2018} and \cite{Scott2018} respectively. 

\subsection{Sample Selection}

The GAMA group sample includes all galaxy associations with two or more members, so to ensure the robustness of our group sample we first select all the SAMI observed galaxies that belong to GAMA haloes with at least 5 members to ensure that they have a robustely estimated velocity dispersion \citep{Robotham2011}. Moreover, only galaxies classified as cluster members by \cite{Owers2017} are taken as cluster galaxies. This gives us an initial sample of 574 galaxies (out of a potential 2502 observed by GAMA) in 205 groups in the GAMA regions and 861 galaxies in 8 clusters. 

Stellar populations measurements are compromised for 100 galaxies because nearby galaxies or stars affect their observation \citep{vandeSande2017}. In addition, we also exclude 3 more galaxies whose $g-i$ colors suggest contamination from other objects in the field of view and 8 galaxies whose annuli do not reflect the shape of their isophotes and/or the central bin is not well aligned with the peak of their flux map (since they could lead to a biased gradient).

\begin{figure}
	\centering
	\includegraphics[height=9cm, width=9cm, keepaspectratio]{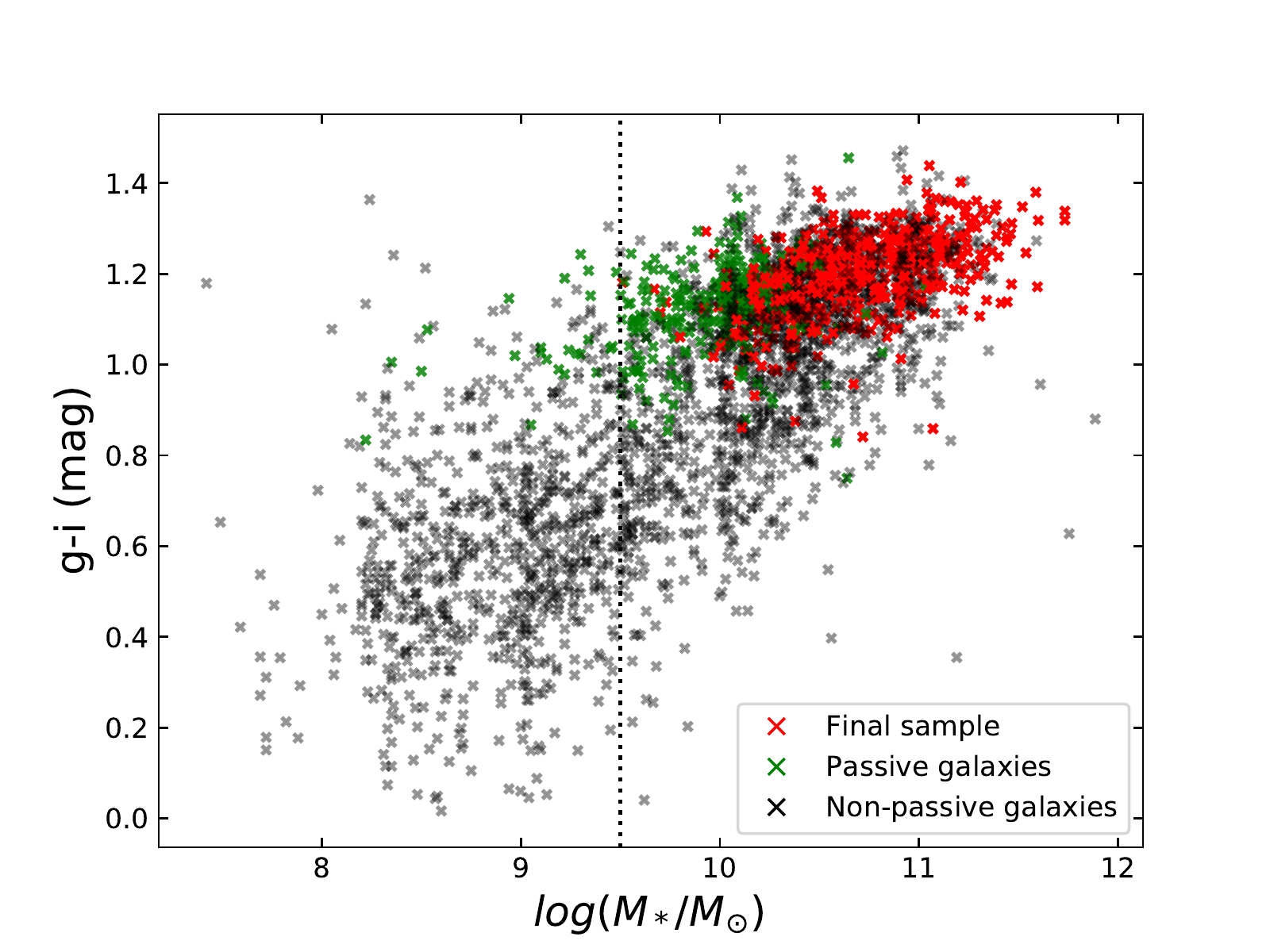} 
	\caption{Distribution of $g - i$ color with stellar mass as a function of galaxy classification for the galaxies included in SAMI v0.11. Unselected passive galaxies are shown as green crosses. Galaxies selected in the final sample are shown as red crosses and non-passive galaxies are shown as black crosses. The dotted line is the stellar mass cut at $\log_{10} (M_*/ M_{\odot}) = 9.5$, galaxies below this mass are excluded from the final sample. }
	\label{fig:color_mass}
\end{figure}

In this paper we are particularly interested in whether there is an environmental dependence to the evolution of central and satellite galaxies. We focus on the passive central galaxies to ensure a like-to-like comparison in determining how their central and satellite status affects their stellar population gradients. We use the SAMI spectroscopic classifications presented in \cite{Owers2019} to select a homogeneous sample. The galaxies are classified as star-forming, passive, or H$\delta$-strong, using the absorption- and emission-line properties of each SAMI spectrum. We select 843 passive galaxies; of these, 98 are central and 745 are satellite galaxies (Fig. \ref{fig:color_mass}).

To ensure the stellar population measurements we use are reliable, we exclude all the galaxies with masses less than $\log_{10} (M_*/ M_{\odot}) = 9.5$, owing to the low S/N and low completeness of these galaxies (we note that selecting galaxies with $\log_{10} (M_*/ M_{\odot}) > 10$ does not change the conclusions we draw). This leaves 819 galaxies (98 central and 721 satellite galaxies). This excludes 40 non-passive central galaxies with $\log_{10} (M_*/ M_{\odot}) > 9.5$. The median halo mass of the non-passive central galaxies is $M_{200} = 10^{13.3} M_{\odot}$.

We also make a quality cut in signal-to-noise ratio and velocity dispersion. For each galaxy, we only use the annular measurements that have S/N $\geq$ 10 and velocity dispersions ($\sigma$) between 75 km/s and 400 km/s. [$\alpha$/Fe] measurements were also restricted to annuli with S/N $\geq$ 20 owing to the greater uncertainties associated with measuring this parameter.

We select galaxies that have at least one annular measurement between 0.1 and 1 $R_e$ and at least 3 in total, all meeting the S/N and $\sigma$ criteria, in order to probe similar spatial regions for our sample of galaxies. This is due to some of the satellite galaxies having effective radii $< 1.1$ - $1.2^{\prime\prime}$, such that the first aperture probes radii greater than 1 $R_e$. 
This radial criterion is the primary source of exclusion for our sample.

These selection criteria result in a final sample of 533 galaxies with metallicity and age gradients (96 central galaxies and 437 satellite galaxies) and 332 galaxies with [$\alpha$/Fe] gradients (79 centrals and 253 satellites).

\begin{figure}
	\centering
	\includegraphics[height=6cm, width=8cm, keepaspectratio]{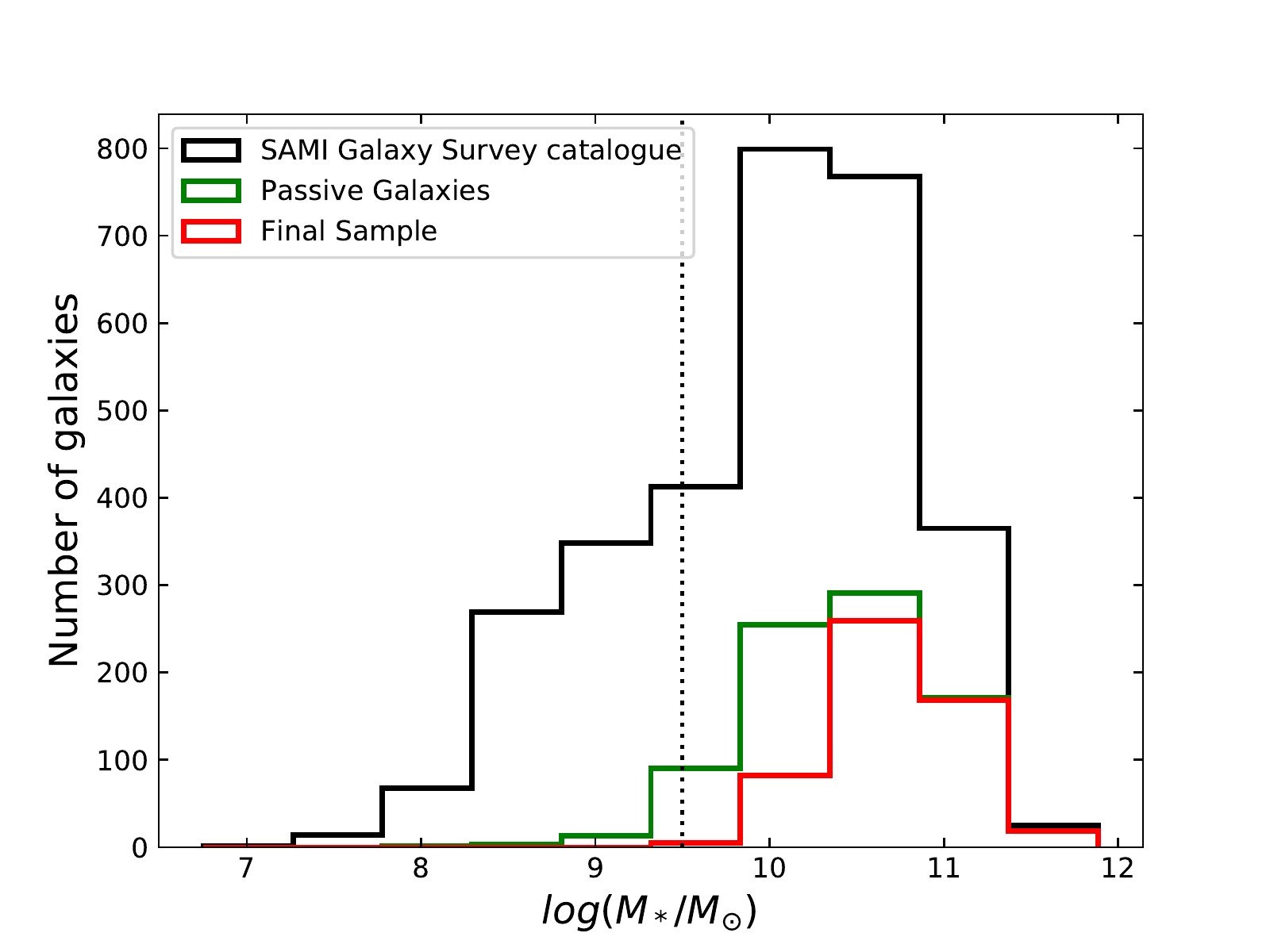}  \includegraphics[height=6cm, width=8cm, keepaspectratio]{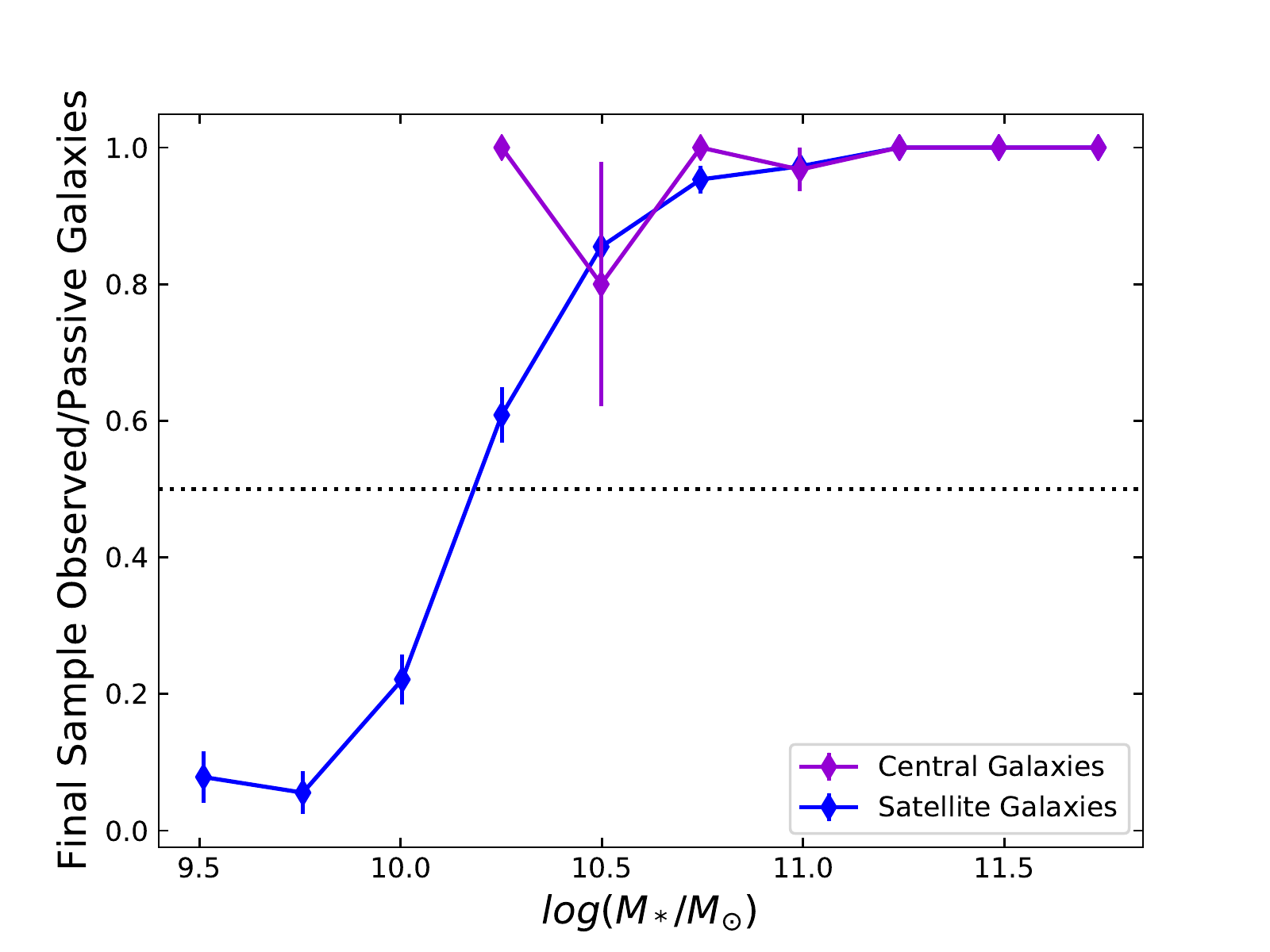} 
	\caption{Sample completeness. Upper panel: the black histogram represents the initial SAMI Galaxy Survey catalogue of 3071 galaxies. Galaxies that are classified as passive are shown in green (843 galaxies). The final sample of central and satellite galaxies is shown in red (533 passive galaxies). The dotted line is the stellar mass completeness cut of $\log_{10} (M_*/ M_{\odot}) = 9.5$. Lower panel: fraction of final galaxies sample selected with respect to the passive galaxy sample for central (purple) and satellite galaxies (blue). Fractional uncertainties are calculated following \cite{Cameron2011}. The final sample is representative in stellar mass. }
	\label{fig:completeness}
\end{figure}

\begin{figure}
	\centering
	\includegraphics[height=6cm, width=8cm, keepaspectratio]{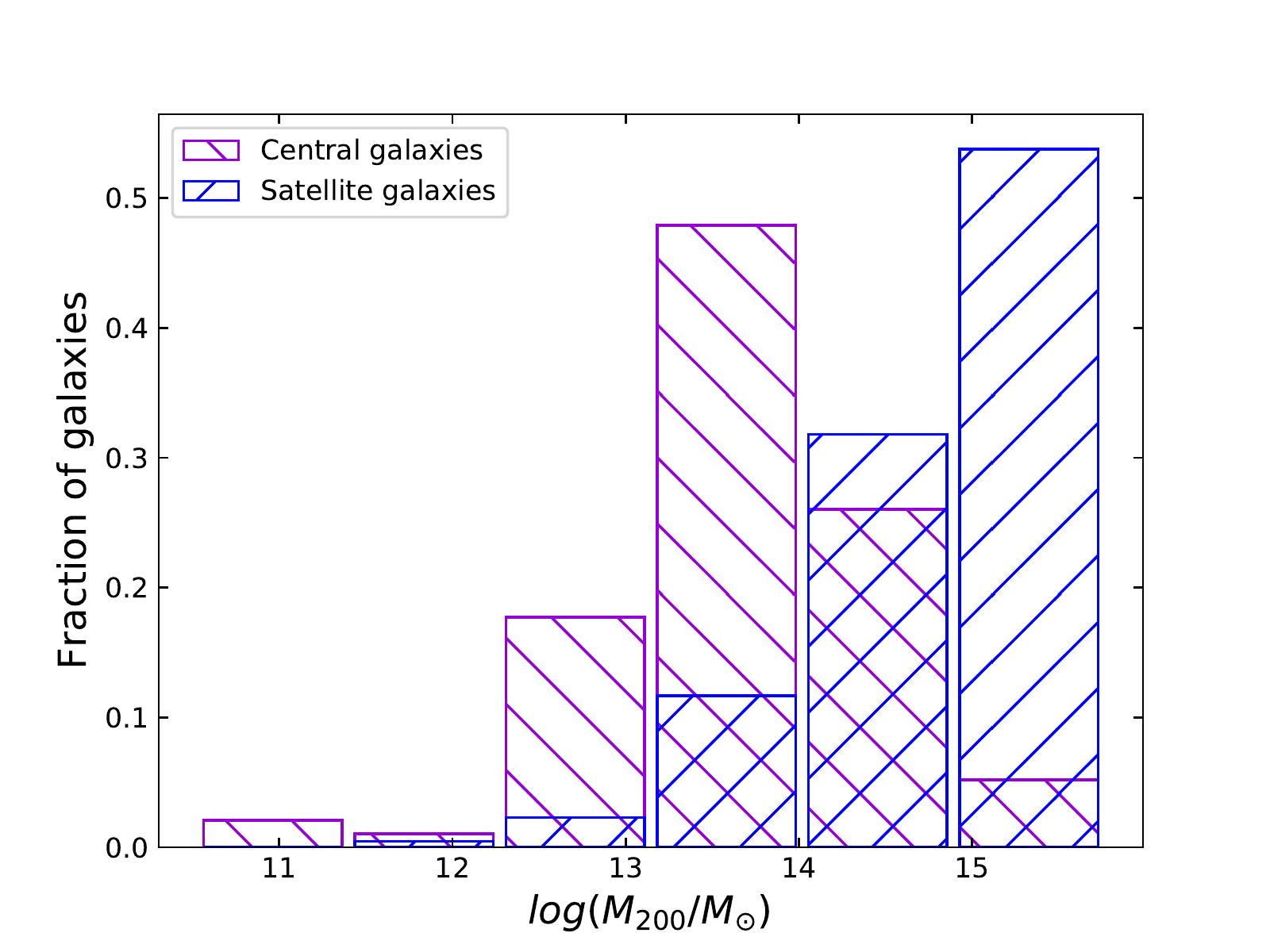}
	\caption{Halo mass distribution of the selected central (purple) and satellite (blue) galaxies. }
	\label{fig:halo_completeness}
\end{figure}

The distribution of $g-i$ color with stellar mass for the galaxies selected is illustrated in Fig. \ref{fig:color_mass}. Fig. \ref{fig:completeness} and Fig. \ref{fig:halo_completeness} illustrate the completeness of the selected sample. As shown, our sample is representative of the parent sample in stellar mass and halo mass, with 96\% completeness for passive galaxies with $\log_{10} (M_*/ M_{\odot}) > 10.5$. 


\subsection{Stellar Mass}
Stellar masses are estimated from $g-$ and $i-$ magnitudes using an empirical proxy developed from GAMA photometry \citep{Taylor2011, Bryant2015}. For cluster galaxies, stellar masses are derived using the same approach \citep{Owers2017}.

The $g-$ and $i-$ magnitudes are taken from Sloan Digital Sky Survey (SDSS; \citealt{York2000}) images for GAMA galaxies and VST/ATLAS (VLT Survey Telescope - ATLAS; \citealt{Shanks2015}) and SDSS DR9 \citep{Ahn2012} observations for cluster galaxies. The VST/ATLAS  data was reprocessed as described in \cite{Owers2017}.

\subsection{Effective Radius}
Circularized effective radii, $R_e$, were determined by fitting a Sersic profile to $r-$band images. For group galaxies, these radii are taken from the GAMA Sersic catalogue v0.7 \citep{Kelvin2012}, whereas for cluster galaxies the $R_e$, expressed as the radius of the major axis, are taken from \cite{Owers2019} and converted to a circularised radius using the axial ratio for each galaxy. For the galaxies that are in common, cluster galaxies' $R_e$ are consistent when measured from VST/ATLAS and SDSS/GAMA \citep{Owers2019}.
\subsection{Halo mass}
We use the GAMA Galaxy Group Catalogue (G3C; \citealt{Robotham2011}) to define the galaxy groups in the GAMA regions. In this catalogue, galaxies are grouped using an adaptive friends-of-friends (FoF) algorithm, taking advantage of the high spectroscopic completeness of the GAMA survey ($\sim 98.5\%$; \citealt{Liske2015}). Following the recommendation of \cite{Robotham2011}, we use the ``iterative center" as the center of our halo. This center is found through an iterative procedure: at each step the $r_{AB}$-band luminosity centroid of the halo is found and the most distant galaxies are rejected. When only two galaxies remain, the brightest is taken as the central galaxy. 

For the GAMA groups we use the Halo Mass, $M_{200}$, the mass contained within $R_{200}$ (the radius at which the density is 200 times the critical density; \citealt{White2001}), calculated from the GAMA group velocity dispersion and calibrated using halos from simulated mock catalogues (see \citealt{Robotham2011}). The cluster halo masses are taken from \cite{Owers2017}, who calculated them from a caustic analysis of the galaxy velocities and their distance from the cluster Centre.

It is important to note that, following \cite{Owers2017}, we apply a scaling factor of 1.25 to the cluster halo masses, to be consistent with the GAMA halo masses. We also scale the GAMA halo masses ($H_0 = 100$ km s$^{-1}$ Mpc$^{-1}$) to be consistent with the cosmology used here. 

\subsection{Central galaxies}
We define the halo central galaxy as the most massive galaxy within 0.25 $R_{200}$ \citep[e.g.,][]{Oliva-Altamirano2017}.
To identify the central galaxies for the group sample, we check that the galaxy identified as ``iterative central" \citep{Robotham2011} is also the most massive galaxy. This is true for 188/205 groups in our sample. 17 groups have different galaxies selected as the iterative central and the most massive. For these 17 groups, we find the most massive galaxy within a radius of 0.25 $R_{200}$. For most of the groups (12/17), the most massive galaxy within 0.25 $R_{200}$ is the iterative central galaxy. For the 5 groups where another galaxy is found to be the most massive within 0.25 $R_{200}$, we select that galaxy as the central galaxy.

A similar procedure is carried out to select the central galaxy in the clusters, in order to ensure consistency between the samples.
We identify the galaxy that sits closest to the center of the cluster (cluster centroids are taken from \citealt{Owers2017}) as well as the most massive galaxy in the cluster.  For 3 out of 8 clusters, these galaxies are the same. For the other 5 clusters, we find the most massive galaxy within a radius of 0.25 $R_{200}$. For 2 clusters this galaxy is also the central one, whereas for 3 out of 5 clusters (Abell 168, Abell 2399 and Abell 4038) the most massive galaxy within 0.25 $R_{200}$ is not the galaxy closest to the center. This is consistent with the dynamical state of these clusters as discussed in \cite{Owers2017} and \cite{Brough2017}. We therefore select the most massive galaxy within 0.25 $R_{200}$ as the central galaxy for these clusters.

\subsection{Data}
We use data from the SAMI v0.11 internal data release (reduced identically to DR2; \citealt{Scott2018}). This data release consists of 3071 galaxies, with repeat observations available for 277 galaxies.\\
3019 SAMI galaxies out of 3071 
have several annular measurements (up to four) of: age, metallicity and $\alpha$-element 
abundance ratio (and their corresponding errors). These are made in elliptical apertures centered on the center of the cube, with constant spacing along the major axis. The position angle, PA, and ellipticity, $\epsilon$, of the galaxy are determined using the \textit{find galaxy} Python routine of \cite{Cappellari2002} from the image generated by summing the cube along its wavelength axis.
The spectra of each of the spaxels in each aperture are summed and the resulting spectrum is analyzed in order to extract the stellar population measurements for each annular aperture.

\subsection{Annular stellar population measurements}
Stellar population measurements were determined as described in \cite{Scott2017}, using an approach based on Lick absorption line strengths. We summarize the method here and identify the differences \cite{Scott2018} take for aperture measurements compared to the total measures presented in \cite{Scott2017}. 

The Lick indices used for this analysis consist of indices defined by \cite{Worthey1997} and \cite{Trager1998}: Balmer lines (H$\delta_A$, H$\delta_F$, H$\gamma_A$, H$\gamma_F$ and H$\beta$); iron-dominated lines (Fe4383, Fe4531, Fe5015, Fe5270, Fe5335 and Fe5406); molecular indices (CN1, CN2, Mg1, Mg2); plus Ca4227, G4300, Ca4455, C4668 and Mg$b$.  All these indices are in the SAMI blue wavelength range.

Before measuring the absorption-line strength in the annular spectra, emission from star formation was removed by comparing the observed spectra with synthetic spectra unaffected by star formation. The best fitting model was found using the penalized Pixel Fitting (pPXF) code of \citealt{Cappellari2004}. This was done by fitting each spectrum with a set of 30 stellar template spectra chosen from the MILES library \citep{Sanchez-Blazquez2006a, Falcon-Barroso2011}. The fit was performed for each input three times: the first run determines the noise in the spectra, the second run identifies bad pixels and emission-affected pixels and the third fit replaces those pixels with the values of the best-fitting emission-free spectra (see \citealt{Scott2017} for details).
This replacement method is more robust than the simple subtraction of emission lines from the spectra, especially in case of low continuum S/N, weak emission or in the wings of emission lines.

In order to measure the Lick indices, all annular spectra were then broadened (by convolving with a Gaussian) in order to match the Lick resolution of the relevant index, also taking into account the combined instrumental broadening, intrinsic broadening due to galaxy's velocity dispersions and additional broadening due to binning over many spaxels.
For every index, a Monte Carlo procedure was used to estimate the uncertainties: noise was randomly added to the best-fitting spectrum and then the indices were remeasured for 100 different realizations and the standard deviation thus found was adopted as the error on each index (see \citealt{Scott2017} for more details).

The indices measured were then converted into single stellar population (SSP) equivalent age, metallicity and $\alpha$-element abundance through comparison with stellar population models that predict Lick indices as a function of logarithmic age, metallicity and [$\alpha$/Fe]. For the SAMI galaxies, two different models were used: \citealt{Schiavon2007} (hereafter S07) and \citealt{Thomas2011} (hereafter TMJ). These models were interpolated to a fine grid (with a resolution of 0.02 in [Z/H] and log age, and 0.01 in [$\alpha$/Fe]). A $\chi^2$ minimization approach (first implemented by \citealt{Proctor2004}) was used to find the SSP that best reproduced the indices measured. As described in \cite{Scott2017}, this also iteratively excluded indices if they were more than 1$\sigma$ outside the range covered by the model. Moreover, the fit would not return a solution if the fitted indices did not include at least one Balmer index and at least one Fe index.
A robust estimation of $\alpha$-element abundance was not always possible due to the S/N of the spectra (S/N $<$ 20). The absorption lines corresponding to the relevant indices were sometimes too shallow or not broad enough to be detected at a high enough S/N.

For each spectrum, the values of age, [Z/H] and [$\alpha$/Fe] were taken from the minimum $\chi^2$ fit and their uncertainties were determined from a $\chi^2$ distribution in the 3D space of the three parameters. 

Comparing the stellar population parameters found with the two models, a good agreement is found for [Z/H] and for [$\alpha$/Fe] (although the latter present a small offset between the two models). The poorest agreement is found comparing ages from the two models, with TMJ predicting older ages. The broad scatter is the result of the fact that the stellar populations are luminosity-weighted and therefore the equivalent mean ages are more sensitive to the youngest populations. The TMJ model predicts older ages at a given H$\beta$, because several galaxies have Balmer line indices that lie outside the TMJ model grid \citep{Kuntschner2010}, therefore the age measurements at low metallicities found by this model are not reliable. After considering all these factors, \cite{Scott2017} adopted the S07 model to derive SSP-equivalent ages, while the TMJ model was used to derive the [Z/H] and [$\alpha$/Fe] values as a best compromise given the limitations, noting that neither model perfectly describes the full set of indices in the SAMI sample. We found that this approach produced good agreement with previous literature studies, whereas the TMJ ages showed an unphysical upturn to much older ages at low masses/metallicities.
However, our work focuses on relative age differences, not absolutes. Considering that relative ages are more accurate, and that the galaxies in the sample presented here do not reach the low metallicities where age reliability is problematic, our results are robust against this issue. \\

\section{Analysis}
\subsection{Derived Gradients}
A log-log linear fit is applied to the annular stellar population measurements of each galaxy, using the python package \textit{scipy.optimize.curve\_fit} \citep{Virtanen2019},
in order to derive the corresponding stellar population gradient. The fit takes into account 
the errors from the stellar population measurements and uses non-linear least squares to fit a straight line to the data.
For the age measurements, since the age uncertainties are asymmetric, we fit the gradient using the mean error on each point.

The fitted slope is taken as the gradient and the uncertainty on the gradient is derived using a Monte Carlo procedure, where each annular value is randomly taken in their uncertainty window and the linear fit is applied again (with the errors taken into account) for 1000 different realizations. The standard deviation from the mean value of the slope is then taken as the uncertainty on the gradient derived.

We note that, as pointed out by \cite{Oyarzun2019}, a non-linear fit could be a better model when deriving the gradients, since the slope of the radial profile can vary going towards the outskirts of the galaxy. However, having only 3 - 4 radial bins for galaxies in our sample, a linear fit with logarithmic radius is the most robust solution. We also note that choosing to fit the gradients up to 1 $R_e$, instead of to the last available radial bin, does not change the conclusions we draw. 

We do not take the inclination of the galaxies into account when deriving the gradients, consistent with previous studies \citep[e.g.,][]{Oliva-Altamirano2015,Greene2015,Goddard2017a,Zheng2017,Ferreras2019}.

\begin{figure}[!ht]
	\centering
	\includegraphics[height=9cm, width=9cm, keepaspectratio]{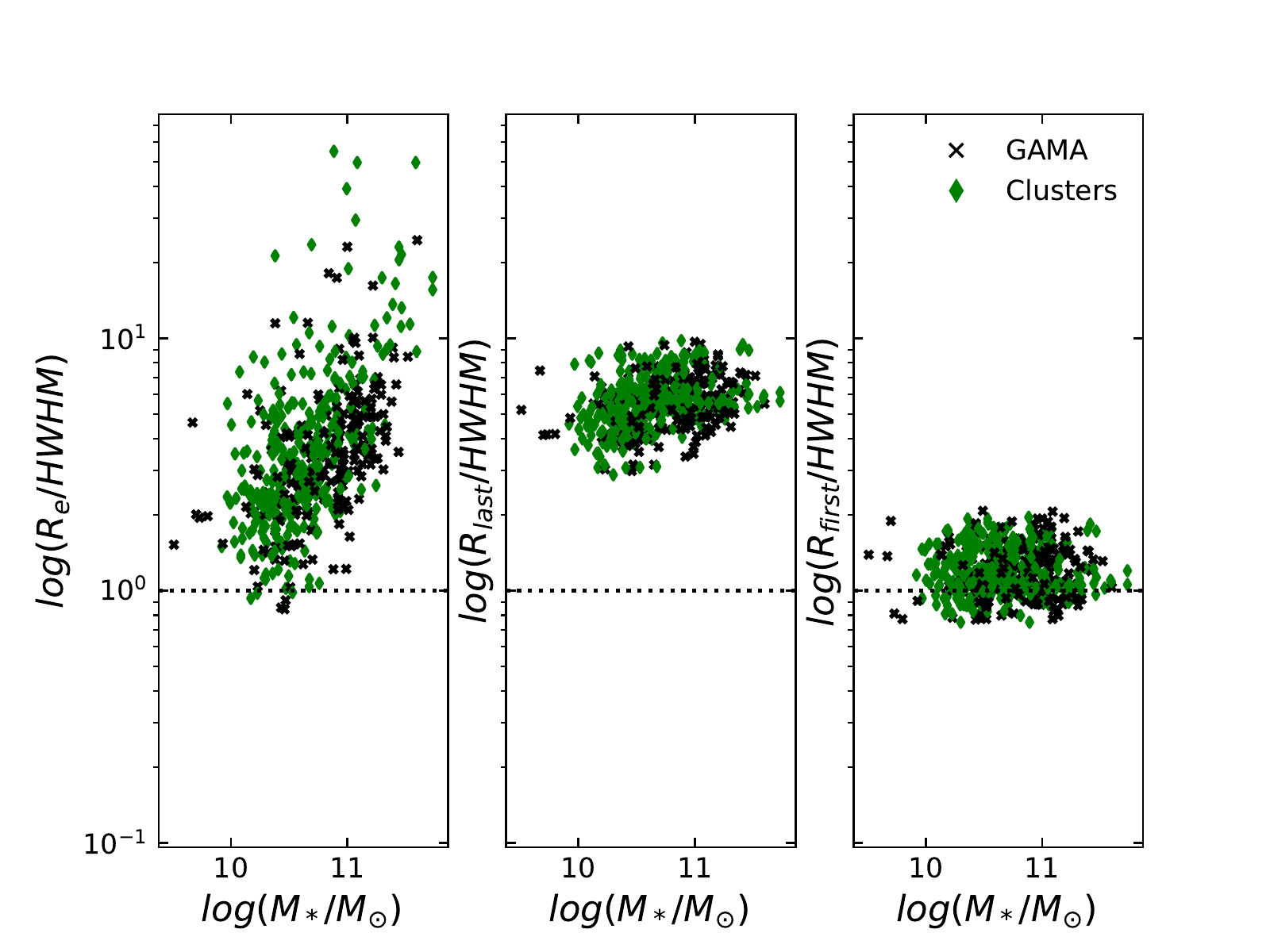} \includegraphics[height=9cm, width=9cm, keepaspectratio]{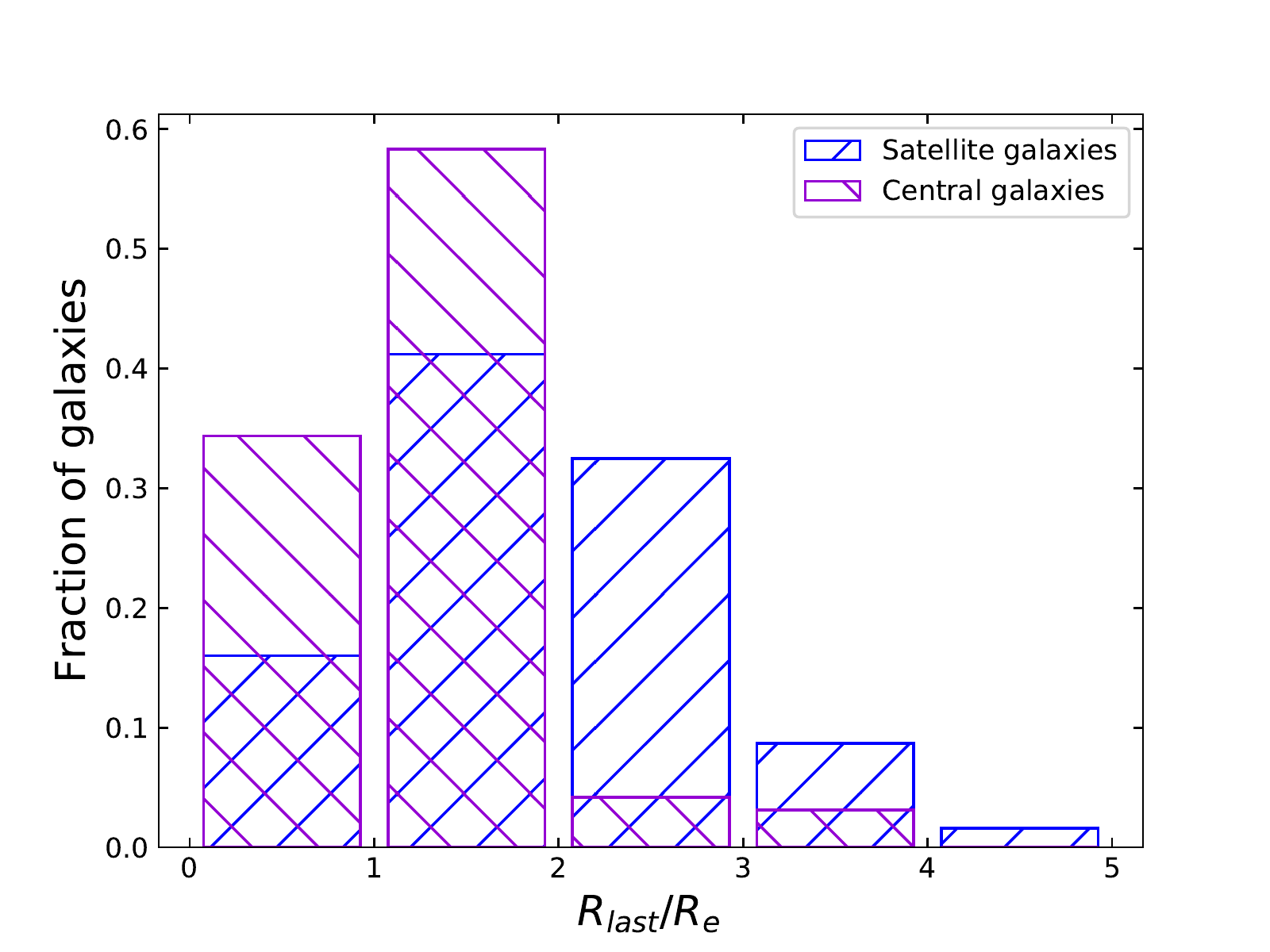} 
	\caption{Upper plot: extent of the spatial resolution of the SAMI observation with respect to the galaxy effective radius, the last radius of the last radial bin and the radius of the first radial bin. Green dots represent cluster galaxies, black crosses are GAMA galaxies. The dotted line shows where the radius value is equal to the Half-Width at Half Maximum (HWHM) of the PSF. Galaxies with values below the dotted line might have gradients affected by the PSF. Lower plot: fraction of central (purple) and satellite (blue) galaxies, over the total number of centrals and satellites, with a given last radial aperture available with respect to the effective radius.
	The majority of central galaxies have stellar population measurements available up to 2$R_e$. }
	\label{fig:psf_re}
\end{figure}

In order to consider the effects of seeing on our results, we compared the Half-Width at Half Maximum (HWHM) of the PSF with the effective radius, the last radial bin and the first radial bin available, as shown in Fig. \ref{fig:psf_re}.
Some of the gradients may be strongly affected by the PSF width, since it is significant compared to the galaxy size (6 galaxies out of 533 have $R_e < $ PSF). 

\begin{figure}[!ht]
	\centering
	\includegraphics[height=9cm, width=9cm, keepaspectratio]{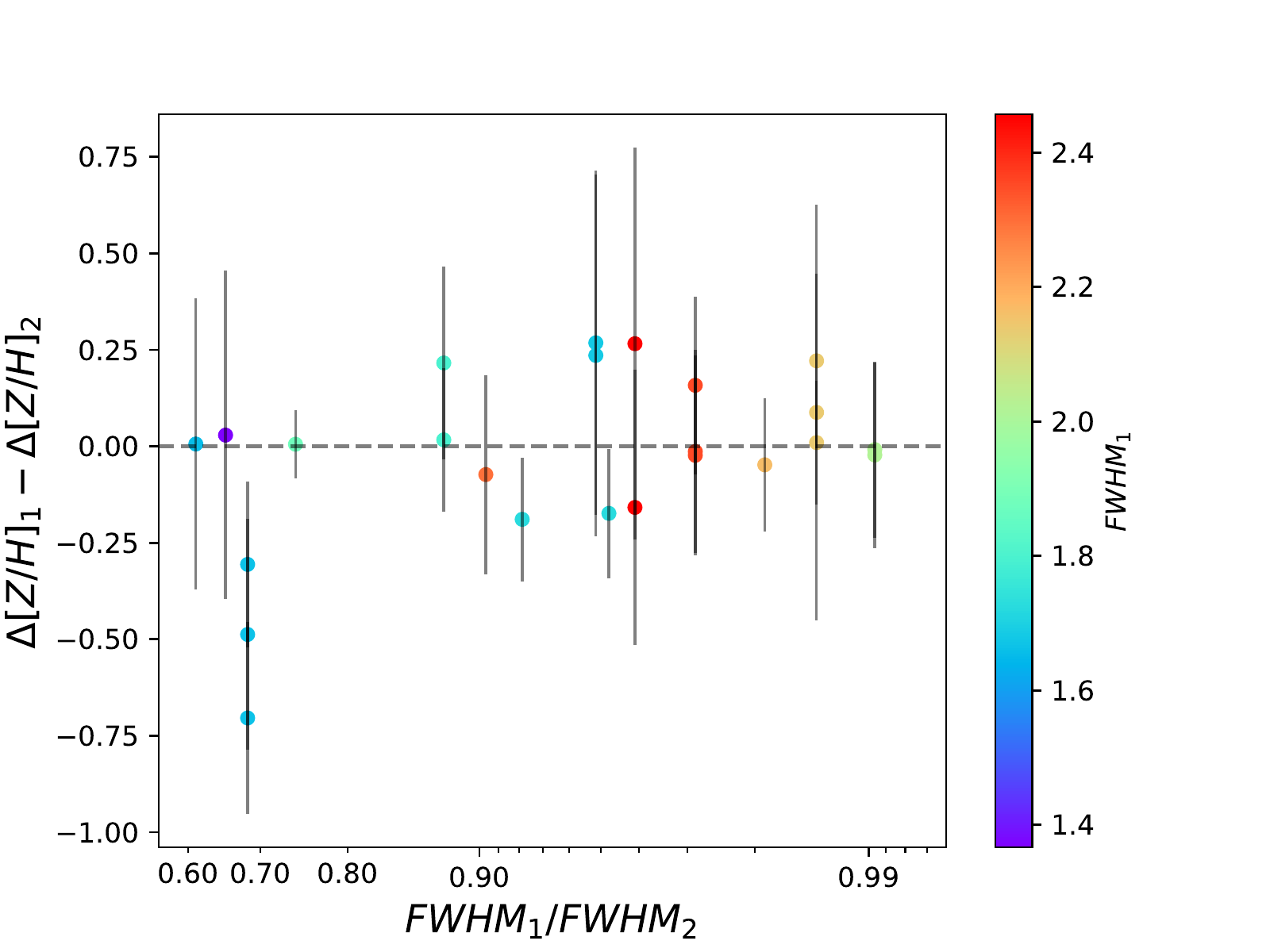} 
	\caption{Difference in the slope of the derived gradient as a function of the difference in seeing for galaxies with repeat observations. The observation with the best seeing is always $Obs_1$. The errors shown are calculated by propagating the errors of the gradients. The points are color-coded according to the values of the best FWHM so that redder points correspond to greater values of $FWHM_1$. The gradients show no significant offset or trend with seeing. }
	\label{fig:psf}
\end{figure}

To measure the effect of the PSF width on the derivation of the gradients, we first analyse the radial profiles of the 23 galaxies in our final sample that have multiple observations available. A small number of the repeat observations exhibit small-scale variations in the flux (aliasing, see \citealt{Green2018} for an extended discussion) that can affect the measured indices, and therefore the derived population parameters. In such cases we identify the discrepant observation by eye, selecting the cube with the more physical spectral variations.
For the remaining 35 galaxies with repeated observations, we measure the slope of the best fit for each observation, finding a mean change in metallicity gradient of $\Delta$[Z/H] = $- 0.03 \pm$ 0.05. Fig. \ref{fig:psf} shows that, within the uncertainties, the measurements are consistent for the majority of the galaxies and show no trend with seeing. However, a more detailed analysis that accounts for the PSF is required to extract robust gradients, with meaningful error bars.

Following \cite{Ferreras2019} we derive the gradients of each galaxy by fitting the observed population parameters in each radial bin to a linear function model convolved with the PSF of the observation. From this model, stellar population values are extracted in the same radial bins used for the observed values and compared with the original observations, applying a standard likelihood function.
The best fit parameters for the slope and the intercept of the linear model are then retrieved (with corresponding uncertainties) using an Markov Chain Monte Carlo sampler (\texttt{EMCEE}, \citealt{Foreman-Mackey2013}).
More details on this process can be found in \cite{Ferreras2019}.

\begin{figure}
	\centering
	\includegraphics[height=6cm, width=8cm, keepaspectratio]{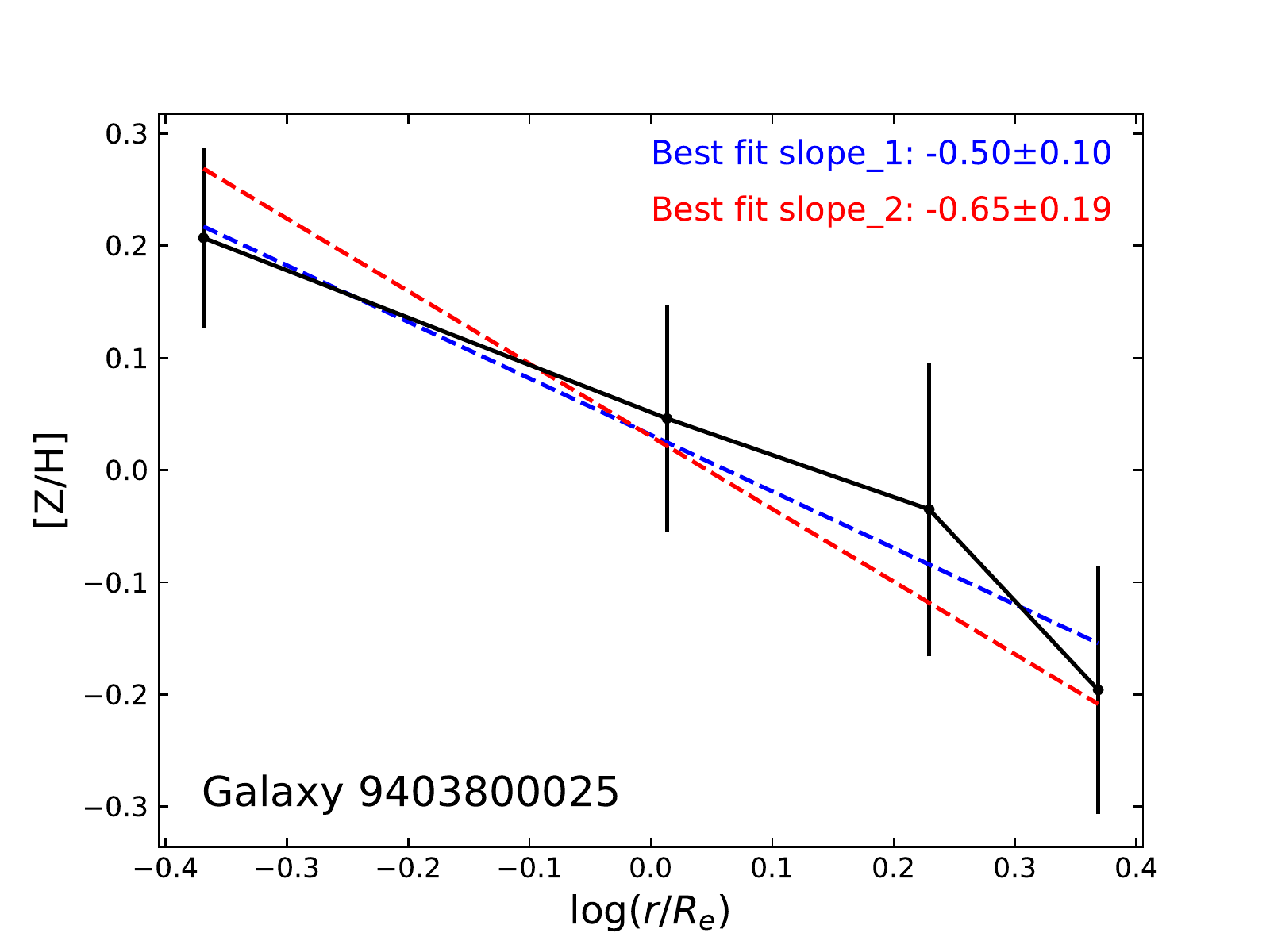}
	\caption{Radial metallicity profile for a typical galaxy. Black points are the metallicity values derived for each aperture, with corresponding uncertainties. The blue dotted line (slope\_1) shows the best-fit to the observed data. The red dotted line (slope\_2) shows the best-fit model after correcting for the effect of the PSF. Best-fit slope values are given in the top right. The best-fit slope derived after correcting is steeper. }
	\label{fig:gal_example}
\end{figure}

The gradients derived following this method are in general steeper than those derived without taking into account the effect of the PSF width, due to the fact that the PSF tends to wash out the gradients.
Fig. \ref{fig:gal_example} shows a radial metallicity profile of a typical galaxy, with the gradient derived using both methods.

Example stellar population maps and radial profiles are shown in Fig. \ref{fig:met_example}, in Appendix B, for 2 example galaxies (1 central and 1 satellite galaxy). 
Hereafter we refer to the gradients derived after correcting for the effect of the PSF. However, considering the gradients derived without taking into account the effect of the PSF does not change the conclusions we draw. 

We explore the consequences Active Galactic Nuclei (AGN) could have on our results in Appendix C, finding that there is no significant influence from AGN emission on our stellar population measurements.

\section{Results}\label{sec:result}
\begin{figure}
	\centering
	\includegraphics[trim=0cm 2.2cm 1cm 3cm,clip=true,height=18cm, width=8cm, keepaspectratio]{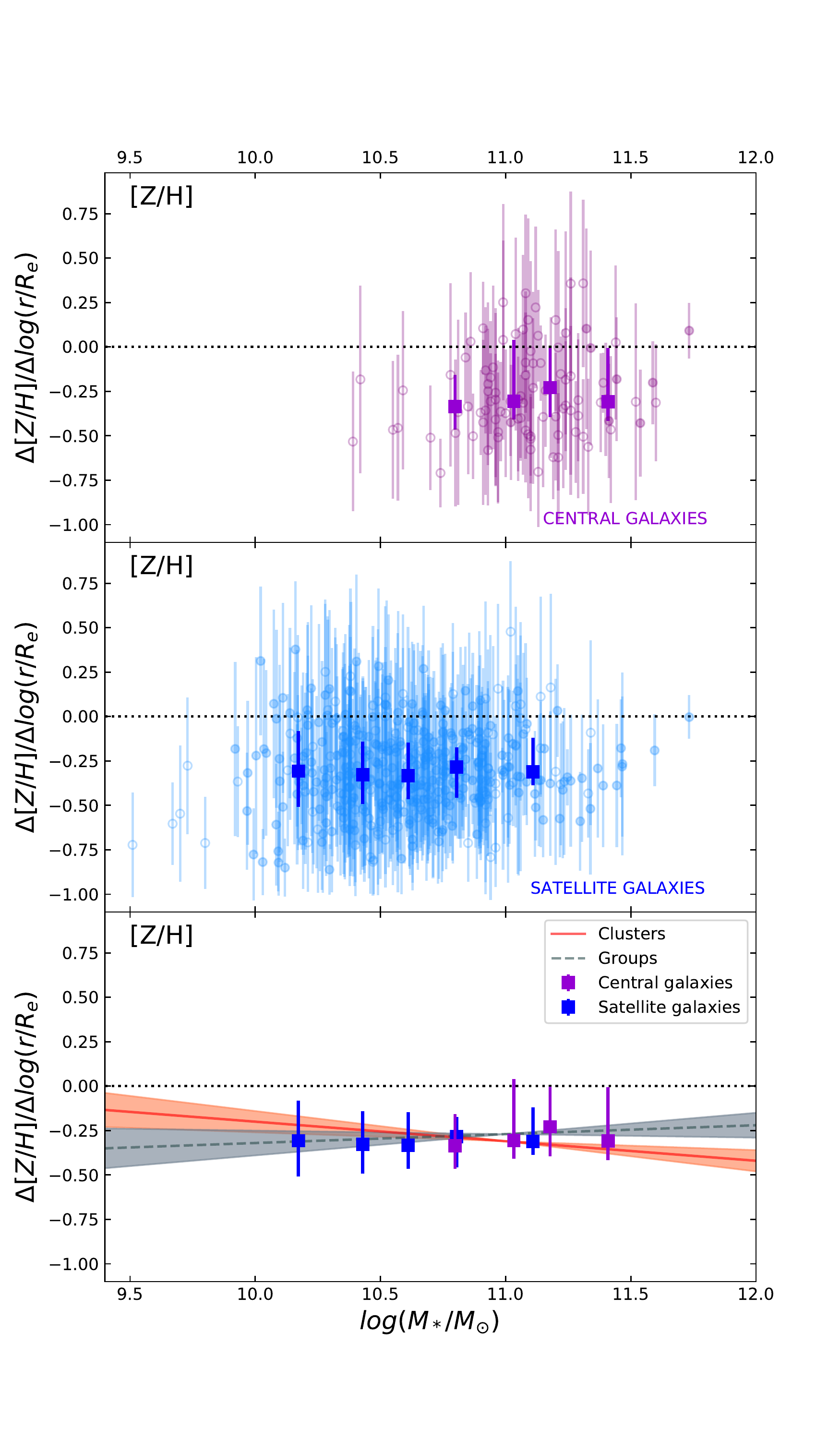}
	\caption{Metallicity gradients as a function of stellar mass. Individual gradients are shown in purple for central galaxies (upper panel) and blue for satellites (middle panel) and the median value in bins of stellar mass in bold.  The lower plot shows the median gradient values for each mass bin in the two environments. Uncertainties shown on the gradient values are calculated as described in Section 3.3 (purple, blue and orange errorbars), while the binned data show the $25^{th}$ and $75^{th}$ percentile in each bin. The dotted line represents a flat gradient. The red and grey lines correspond to the gradients found for cluster and group galaxies, respectively, by \cite{Ferreras2019}. Metallicity gradients are negative and shallow, for both central and satellite galaxies. }
	\label{fig:met}
\end{figure}

\begin{figure}
	\centering
	\includegraphics[trim=0cm 2.2cm 1cm 3cm,clip=true,height=18cm, width=8cm, keepaspectratio]{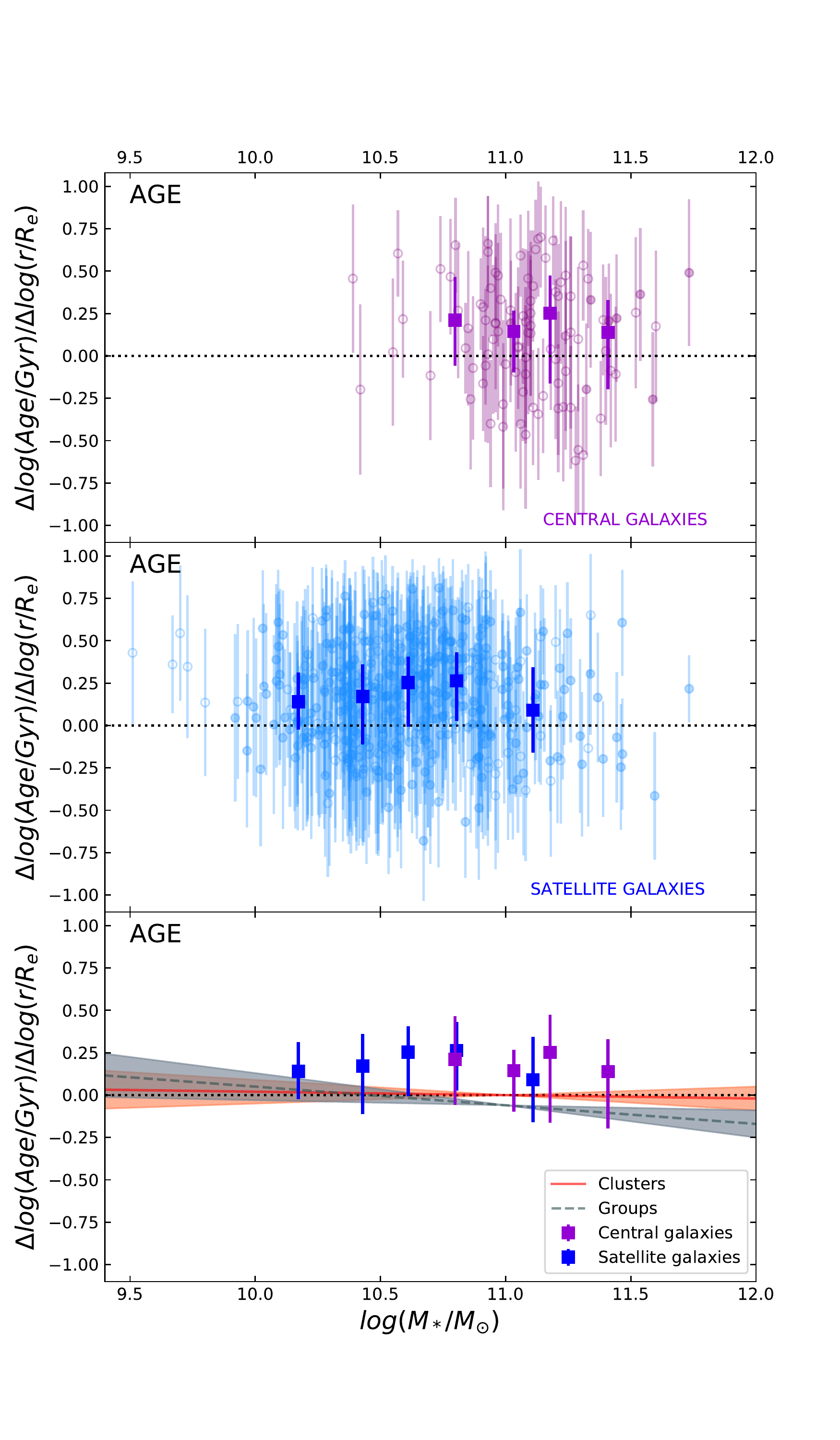}
	\caption{Age gradients as a function of stellar mass. All point types are as per Fig. \ref{fig:met}. Age gradients are generally positive and shallow for both central and satellite galaxies.}
	\label{fig:age}
\end{figure}

\begin{figure}
	\centering
	\includegraphics[trim=0cm 2.2cm 1cm 3cm,clip=true,height=18cm, width=8cm, keepaspectratio]{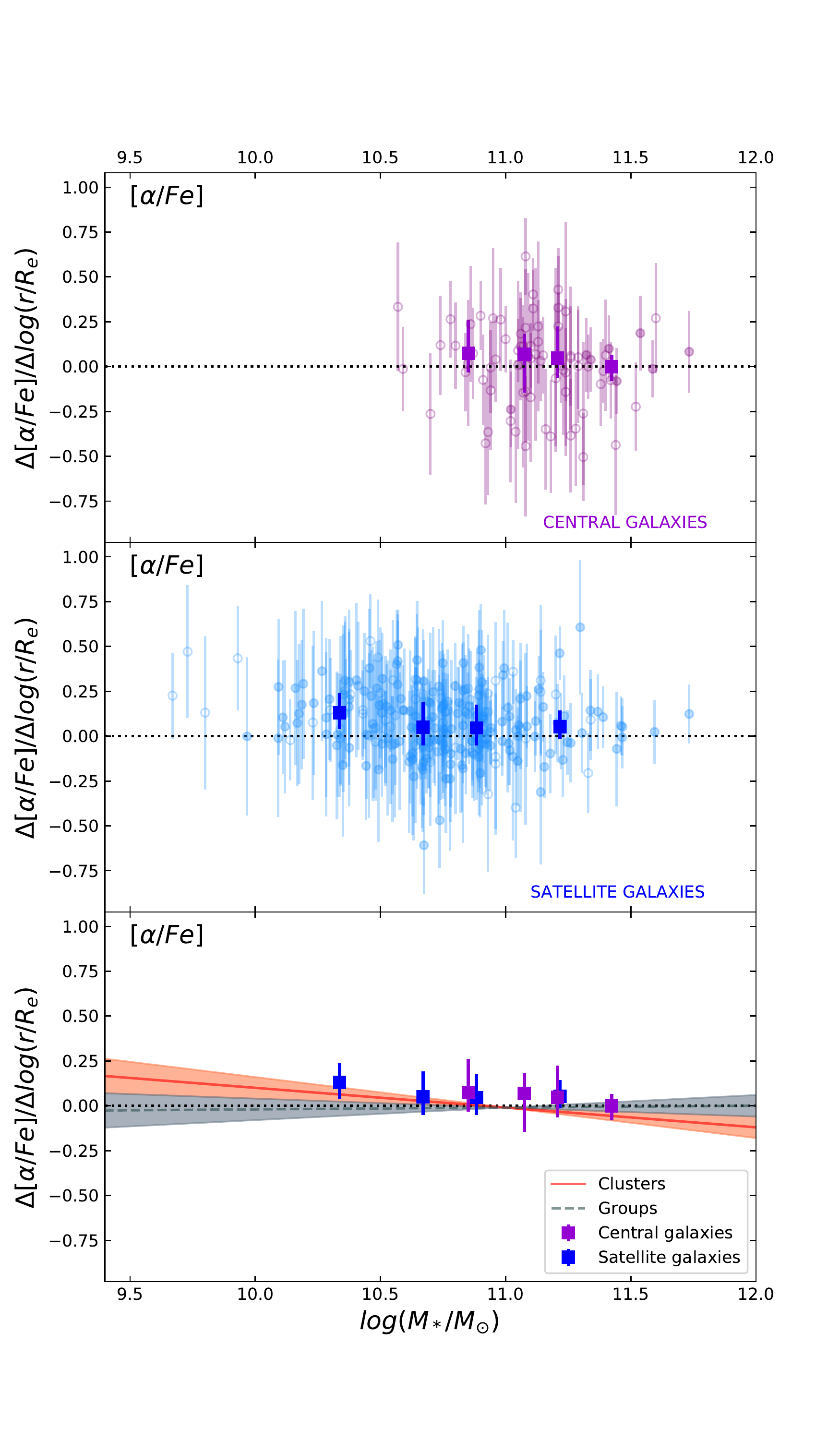}
	\caption{[$\alpha$/Fe] gradients as a function of stellar mass. All point types are as per Fig. \ref{fig:met}. [$\alpha$/Fe] gradients are consistent with zero for both central and satellite galaxies. We find a weak trend of the gradients with stellar mass. }
	\label{fig:alpha}
\end{figure}

In this section we present the results we obtain for the stellar population gradients of passive central and satellite galaxies in the SAMI Galaxy Survey. Fig. \ref{fig:met} to \ref{fig:alpha} show the binned gradients (metallicity, log age and $\alpha$-element 
abundance ratio) as a function of stellar mass.
To better understand and visualise any trends underlying these gradients,
we group the gradients together by stellar mass in bins of 25 galaxies each for the central galaxies and 90 for the satellite galaxies, in the plots showing age and metallicity gradients. We use bins of 20 galaxies for centrals, and of 70 galaxies for satellites for [$\alpha$/Fe] gradients. 

Figure \ref{fig:met} shows that the metallicity gradients are negative and generally shallow ($\Delta$[Z/H]/$\Delta \log(r/R_e)\gtrsim$ $-$0.3), at all masses. To test whether there is any statistically significant trend with mass for the whole sample, we use the Kendall's correlation coefficient $\tau$, using the Python package \textit{scipy.stats.kendalltau} \citep{Virtanen2019}. This correlation coefficient is robust to small sample sizes. A $\tau$ value close to 1 indicates strong agreement, while a value close to $-$1 indicates strong disagreement. We find a value of $\tau = 0.05$, with a probability of correlation of 93.8\%. This suggests the possibility of a weak trend that the gradients tend to be shallower (less negative) with increasing mass. 

Figure \ref{fig:age} shows that age gradients are slightly positive, showing no trend with stellar mass ($\tau = -0.02$, with a probability of correlation of 60\%) and no difference in trend between central and satellite galaxies.\\
Figure \ref{fig:alpha} shows that $[\alpha/Fe]$ gradients are consistent with zero and show a weak negative trend with stellar mass ($\tau = -0.13$, with a probability of correlation of 99.96\%) and no difference between the different environments.\\
The mean gradient values for central and satellite galaxies are given in Table \ref{tab:mean_values}. The gradients for central and satellite galaxies are consistent within 1$\sigma$.

\begin{table*}[!ht]
	\begin{center}
		\begin{tabular}{ |c|c|c| } 
			\hline & \textbf{Central galaxies}  (1$\sigma$)& \textbf{Satellite galaxies} (1$\sigma$)\\
			 \hline
			 $\Delta [Z/H]/ \Delta \log(r/R_e)$ & $-0.25 \pm 0.02$ (0.24) & $-0.30 \pm 0.01$ (0.24)\\
			 $\Delta (\log Age)/ \Delta \log(r/R_e)$ & $0.13 \pm 0.03$ (0.33) & $0.17 \pm 0.01$ (0.29)\\
			 $\Delta [\alpha/Fe]/ \Delta \log(r/R_e)$ & $0.01 \pm 0.03$ (0.23)& $0.08 \pm 0.01$(0.18)\\  
			\hline
			
		\end{tabular}
	\end{center}
	\caption{Mean $[Z/H]$, $(\log Age)$ and $[\alpha/Fe]$ gradients for central and satellite galaxies. Uncertainties shown are errors on the mean. The 1$\sigma$ deviations are shown in parenthesis.}\
	\label{tab:mean_values}
\end{table*}

In order to test whether the scatter in the observed relations is due to measurement errors or an underlying intrinsic scatter, we use the \texttt{LtsFit} Python package, implementing the method used in \cite{Cappellari2013}. This package performs a linear regression which allows for intrinsic scatter and observational errors in all coordinates. We applied this test to the subsets of central and satellite galaxies separately and to the whole sample. We found no evidence of a statistically significant intrinsic scatter for any of the observed relations.

\subsection{Host halo mass dependence}

In order to determine whether there is any dependence of the gradients on host halo mass, we split our central galaxy sample into three subsets of halo mass. Each subset has an equal number of galaxies (32 galaxies for metallicity and age gradients and 26 galaxies for [$\alpha$/Fe] gradients). We then bin the satellite galaxies into the same halo mass subsets as the central galaxies.

\begin{figure}
	\centering
	\includegraphics[trim=0cm 2.2cm 1cm 3cm,clip=true,height=18cm, width=8cm, keepaspectratio]{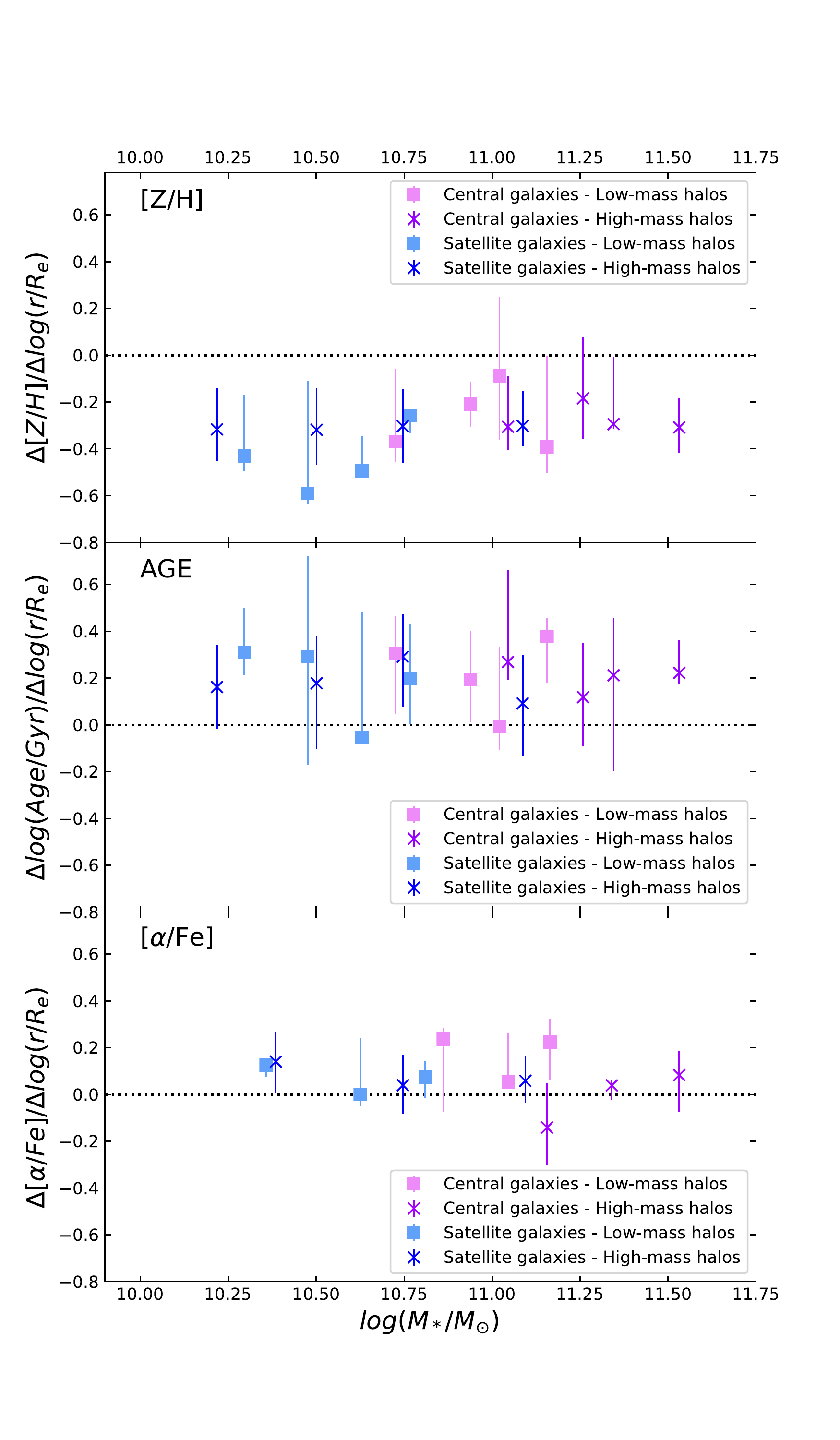}
	\caption{Median values of the metallicity gradients as a function of stellar mass and halo mass. Gradients for central galaxies are shown as pink squares (low-mass halos), and purple crosses (high-mass halos). Gradients for satellites are shown in light blue squares (in the low-mass halos) and dark blue crosses (high-mass halos). Uncertainties shown on the binned gradient values are the $25^{th}$ and $75^{th}$ percentile in each bin. The dotted line represents a flat gradient. Metallicity gradients for central and satellite galaxies are negative, and consistent within the errors for different halo masses. The metallicity gradients of galaxies in low-mass halos (both satellite and central galaxies) become shallower with increasing stellar mass compared to the gradients of galaxies in denser environments, which do not show any trend with stellar mass. The age gradients are generally positive for both central and satellite galaxies, with no differences between low-mass and high-masss halos. [$\alpha$/Fe] gradients are generally consistent with zero for both central and satellite galaxies and show no dependence on host halo mass.}
	\label{fig:grad_env}
\end{figure}

Figure \ref{fig:grad_env} shows the stellar population gradients of central and satellite galaxies as a function of stellar mass, in the low-halo mass (10.96 $< M_{200} <$ 13.28) and high-halo mass (13.77 $< M_{200} <$ 15.29) subsets. Each subset has been grouped into bins by stellar mass (similar to the mass bins described at the beginning of Section \ref{sec:result}) to better visualize any underlying trend with stellar mass.

The metallicity gradients of central and satellite galaxies show no differences between those in low-mass halos and high-mass halos.
In the stellar mass bins where we have values for both central and satellite galaxies, the metallicity gradients of central galaxies are consistent with those of satellite galaxies in structures of similar halo mass. The metallicity gradients of galaxies in low-mass halos (both satellite and central galaxies) become shallower with increasing stellar mass compared to the gradients of galaxies in denser environments, which do not show any trend with stellar mass. Using Kendall's correlation coefficient, we find a value of $\tau = 0.24$, with a probability of correlation of 98.9\%. This weak trend could be the driver of the weak trend with stellar mass that we see for the whole sample of galaxies.

Age gradients are generally positive for both central and satellite galaxies, with no trend with stellar mass regardless of the host halo mass.

Central and satellite galaxies show similar [$\alpha$/Fe] gradients for both environments.

\section{Discussion}
Integral Field Spectroscopy (IFS) has been used as a valuable tool to explore the
stellar population parameters of galaxies, due to its ability to map stellar populations across galaxies. Here we have analysed the stellar population gradients of 533 passive galaxies from the SAMI Galaxy Survey. Of these 96 are central galaxies and 437 are satellite galaxies.

\subsection{Metallicity gradients}
We observe shallow metallicity gradients in our sample (mean $\Delta[Z/H]/\Delta \log(r/R_e)$ = $-$0.25 $\pm$ 0.02, with a 1$\sigma$ deviation of 0.24, for central galaxies and for satellite galaxies a mean $\Delta[Z/H]/\Delta \log(r/R_e)$ = $-$0.30 $\pm$ 0.01, with a 1$\sigma$ deviation of 0.24). These are in agreement with previous results from \cite{Kuntschner2010} and \cite{Oliva-Altamirano2015}.

Examining the relationship of metallicity gradients with stellar mass (Fig. \ref{fig:met}), we find a potential weak positive trend such that gradients are shallower (less negative) with increasing stellar mass. This is in agreement with \cite{Zheng2017}, who found evidence of a weak positive trend of metallicity gradients with stellar mass, so that the gradients become less negative with increasing stellar mass, in a sample of 577 early-type MaNGA galaxies. Their trend is more evident when examining galaxies in low-mass halos (sheet and void environment). 

We do not observe, however, a transition to a shallower metallicity gradient at $\log_{10} (M_*/ M_{\odot})= 10.54$ as observed by \cite{Kuntschner2010} and predicted by \cite{Taylor2017} from chemodynamical simulations.

Our lack of significant correlation with stellar mass is in contrast to \cite{Li2018}, who found a clear decrease of the metallicity gradients (i.e. more positive gradients) with increasing stellar mass, and \cite{Goddard2017a}, who found a weak negative correlation of gradients with stellar mass in a sample of MaNGA ETGs. However, we note that if we examine the gradients derived by \cite{Goddard2017a} and \cite{Li2018} for galaxies in the same mass range as those in our sample ($9.5 < \log_{10} (M_*/ M_{\odot}) < 11.8$), they show no strong dependence with stellar mass, in agreement with our results.

\cite{Tortora2012}, in a sample of about 4500 SDSS early-type galaxies, found that age and metallicity gradients generally do not depend on environmental density, in agreement with our results. However, they find a residual dependence of metallicity gradient with environment for central galaxies (shallower metallicity gradients in denser environments).

\cite{Ferreras2019} made independent stellar population measurements for 522 early-type galaxies in the SAMI Galaxy Survey and found a marginal steepening of metallicity gradient with increasing stellar mass. They found the steepening to be stronger in the denser cluster environments, whereas the group environments appear to have shallower metallicity gradients.
While we do not see a steepening in the metallicity gradients in denser environments, \cite{Ferreras2019} mean gradients are consistent with our derived gradients, as shown in Fig. \ref{fig:met} (lower panel) and the relationships are consistent within their errors. The difference in the metallicity gradient-stellar mass relationship observed by \cite{Ferreras2019} could therefore be driven by the different selection of halo structures.

\begin{figure}
	\centering
	\includegraphics[height=18cm, width=8cm, keepaspectratio]{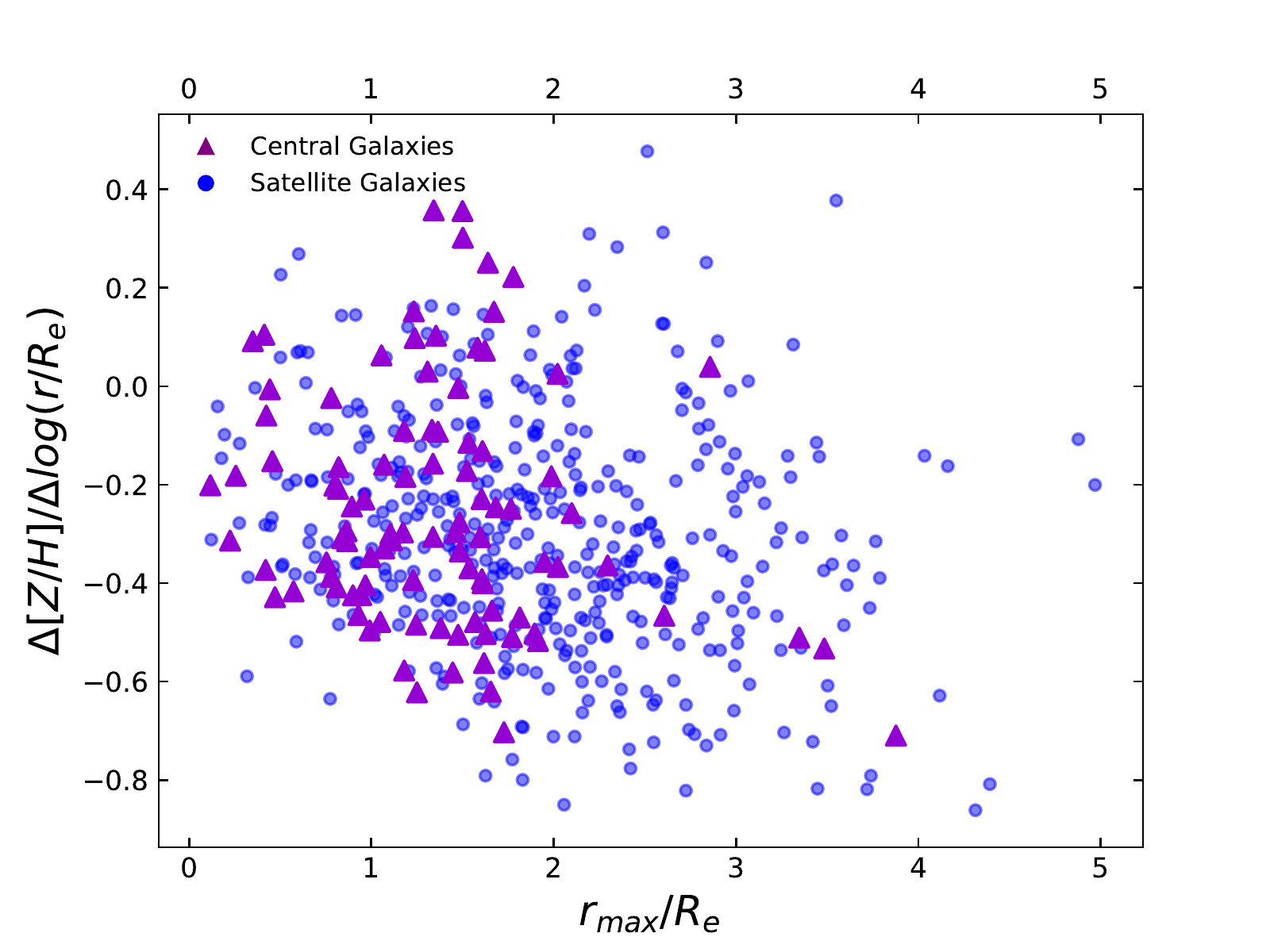}
	\caption{[Z/H] gradients as a function of the last radial bin available for each galaxy (in units of $R_e$). Purple triangles represent central galaxies and blue dots represent satellite galaxies. For central galaxies, the majority of the gradients are derived within 2 $R_e$.}
	\label{fig:grad_r}
\end{figure}

\subsection{Age gradients}
We find mildly positive age gradients (i.e. younger centers), with mean $\Delta \textrm{log (Age/Gyr)}/\Delta \log(r/R_e) =0.13 \pm 0.03$ for central galaxies and for satellite galaxies mean $\Delta \textrm{log (Age/Gyr)}/\Delta \log(r/R_e) =0.17 \pm 0.01$, with 1$\sigma$ deviations of 0.33 and 0.29 respectively.

Evidence of flat or slightly positive gradients in age have been previously observed by numerous studies, such as \cite{Sanchez-Blazquez2007, Brough2007, Kuntschner2010, Loubser2012, Greene2015, Goddard2017a}.

We find no correlation of age gradients with stellar mass, in agreement with \cite{Zheng2017}.

\cite{Ferreras2019} found that there is a slight steepening of the age gradients towards high-mass galaxies (with a slope of -0.07 $\pm$ 0.04), in particular galaxies in groups appear to have steeper age gradients (i.e. more negative) than galaxies in clusters, as shown by the red and grey lines in Fig. \ref{fig:age} (lower panel). When examining our sample as a function of halo mass, we do not see any trend of age gradients with stellar mass, nor any dependence on halo mass (in agreement with \citealt{Tortora2012}). While our relationships are not consistent with \cite{Ferreras2019}, their mean gradients are consistent with our derived gradients, as shown in Fig. \ref{fig:age}. The difference in relationship is likely driven by the different selection of low- and high-mass halos for the analysis (\citealt{Ferreras2019} divided galaxies into group vs cluster galaxies).

\subsection{[$\alpha$/Fe] gradients}

We find flat and shallow [$\alpha$/Fe] gradients, with a mean value of $\Delta \log [\alpha/Fe]/\Delta \log(r/R_e) = 0.01 \pm 0.03$ for central galaxies, and for satellite galaxies $\Delta \log [\alpha/Fe]/\Delta \log(r/R_e) = 0.08 \pm 0.01$, with standard deviations of 0.23 and 0.18 respectively. These results are in good agreement with \cite{Kuntschner2010}, who found $\Delta \log [\alpha/Fe]/\Delta \log(r/R_e)$ between 0.01 and 0.07.

We find a weak negative correlation of [$\alpha$/Fe] gradients with stellar mass, such that the gradients are more negative with increasing masses (Fig. \ref{fig:alpha}).
\cite{Greene2015}, with a sample of 50 MASSIVE galaxies and 50 lower-mass galaxies, found similar [$\alpha$/Fe] gradients, but their correlation is opposite to that found here: galaxies in their low-mass bin ($10.1 \gtrsim \log_{10} (M_*/ M_{\odot}) \lesssim 11.2$) have $\Delta \log [\alpha/Fe]/\Delta \log(r/R_e) = - 0.09 \pm 0.06 $ and  $\Delta \log [\alpha/Fe]/\Delta \log(r/R_e) = 0.1 \pm 0.08 $ in the higher-mass bin ($\log_{10} (M_*/ M_{\odot})> $ 11.6). However, the gradients are consistent within our 1$\sigma$ deviation.

\cite{Greene2019} found that [$\alpha$/Fe] gradients for very high-mass galaxies (11.6 $< \log_{10} (M_*/ M_{\odot})< $ 12.2) are consistent with 0, with a median value of $\Delta \log [\alpha/Fe]/\Delta \log(r/R_e) = - 0.03 \pm 0.02 $, in agreement with our median value for central galaxies $\Delta \log [\alpha/Fe]/\Delta \log(r/R_e) = 0.01 \pm 0.03 $.

\cite{Ferreras2019} found evidence of [$\alpha$/Fe] gradients varying from positive to more negative with increasing mass for cluster galaxies (Fig. \ref{fig:alpha}), in agreement with what we see for central galaxies in high-mass halos. However, we do not see, more negative [$\alpha$/Fe] gradients for satellite galaxies in high-mass halos.

Similarly shallow positive [$\alpha$/Fe] gradients are also found in simulations, \cite{Hirschmann2015} derived mildly positive age gradients ($\Delta \textrm{log (Age/Gyr)}/\Delta \log(r/R_e) \approx 0.04$). \cite{Taylor2017} found small positive [$\alpha$/Fe] gradients ($\Delta \log [O/Fe]/\Delta \log(r/R_e) \approx 0.1-0.2$).

\subsection{Implications}\label{sec:mergers}  
In this paper we find negative metallicity gradients, positive age gradients and flat [$\alpha$/Fe] gradients. The metallicity gradients show evidence of a potential weak trend with stellar mass, so that gradients become less negative with increasing stellar mass, while [$\alpha$/Fe] gradients become more negative with increasing mass. When  examining the galaxies as a function of halo mass, we find evidence of shallower (less negative) metallicity gradients with increasing stellar mass for galaxies in low-mass halos, but no significant difference between the gradients of central and satellite galaxies. 

The gradients we find are consistent with those predicted for galaxies formed in a two-phase process \citep{White1980,Kobayashi2004, Oser2010}. In this scenario, galaxies first undergo a rapid formation phase during which ’in-situ’ stars are formed within the galaxy. The more massive ETGs start forming stars at earlier epochs \citep{DeLucia2006, Dekel2009}. The resulting galaxy is compact, with a steep negative metallicity gradient (with gradients ranging from
$\Delta[Z/H]/\Delta \log (r/R_e)$ = $-$0.35 to $-$1.0; \citealt{Kobayashi2004}) and a positive age gradient (due to the presence of a younger stellar population in the center of the galaxy). More specifically, the metallicity and age gradients would naturally correlate with galaxy mass as star formation lasts longer in the center of more massive systems which have deeper central potential wells (e.g. \citealt{Kobayashi2004, Pipino2010}), so that more massive galaxies would have younger and more metal-rich centers, more positive age gradients and more negative metallicity gradients.

After the star formation is quenched, a second phase of slower evolution is dominated by dissipationless mergers.
In particular, galaxy interactions will either steepen or flatten the metallicity gradients: major dissipationless mergers tend to cause gradients to become shallower, as metal-rich stars from the centre of the galaxy are redistributed further out \citep{Kobayashi2004, DiMatteo2009, Rupke2010, Navarro-Gonzalez2013, Taylor2017} and minor dissipationless mergers lead to steeper negative metallicity gradients, due to metal-poor stars being accreted in the outer regions from lower mass galaxies \citep{Hilz2012, Hilz2013,Hirschmann2015}.

Central galaxies, due to their privileged position in the center of the potential well, are expected to experience a greater number of galaxy interactions compared to satellite galaxies. 
In this scenario, simulations predict flat or slightly positive age gradients, due to old stars being added in the outer regions (at radii $> 2 R_e$; \citealt{Hirschmann2015}), and [$\alpha$/Fe] profiles are expected to be flat as a result of massive early-type galaxies assembling via mergers with low-mass systems \citep{Gu2018}.
These predictions are in agreement with our findings: the metallicity gradients we find (see Table \ref{tab:mean_values}) are shallower than those predicted by a monolithic collapse (ranging from $\Delta[Z/H]/\Delta \log (r/R_e) = -0.5$ \citealt{Carlberg1985} to $\Delta[Z/H]/\Delta \log (r/R_e) =-1.0$ \citealt{Kobayashi2004}), hinting to a flattening of the gradients due to mergers. The age gradients are slightly positive (with central stars having younger ages than stars in the outskirts) and generally flat [$\alpha$/Fe] gradients. However, since evidence of  minor later accretions are expected at radii $>$ 1.5-2$R_e$ and the majority of the radial profiles for our central galaxies only extend to a maximum of 2$R_e$ (as shown in Fig. \ref{fig:grad_r}), our results point to a similar evolution path for the inner regions of both central and satellite galaxies.

Previous works have seen clear evidence that satellite galaxies in these environments are found to be older than central galaxies of the same stellar mass in higher density environments \citep{LaBarbera2014, Scott2017,Thomas2005,Pasquali2010}. However we do not see a difference in the gradients of central and satellite galaxies when taking into account the mass of their host halos.

The fact that our metallicity gradients for central and satellite galaxies are consistent within the standard deviation suggests that the inner regions of central galaxies (up to 2 $R_e$) form in a similar fashion to similarly massive satellite galaxies. Moreover, the gradients for galaxies in high-mass halos show no trend with stellar mass. This seems to point to a similar evolution for galaxies in denser environments, regardless of their current position in the halo.

The potential trend of shallower metallicity gradients with increasing stellar mass in low-mass halos could point to a greater number of interactions in the group environment. Galaxy interactions in this environment are more likely due to the lower relative velocity at which galaxies are moving. In high-mass halos the frequency of galaxy mergers is low, due to the fast orbital motions of the galaxies and the long time scales for dynamical friction \citep[e.g.][]{Park2009}. 

The consistent stellar population gradients ([Z/H], age and [$\alpha$/Fe]) presented here for both central and satellite galaxies suggest a similar formation and evolution history for the inner regions (R $<$ 2$R_e$) of central and satellite galaxies, with no dependence on their host halo mass.

\section{Conclusions}
We have analysed the stellar population radial profiles obtained from IFU observations of a large sample of passive galaxies (with $9.5 < \log_{10} (M_*/ M_{\odot}) < 11.8$), spanning different environments, in the SAMI Galaxy Survey. We have compared the derived gradients of central galaxies to those of satellite galaxies. Our final sample of passive galaxies consists of 533 galaxies including 96 central and 437 satellite galaxies. The sample used for [$\alpha$/Fe] analysis is smaller (332 galaxies), due to the higher S/N required. We draw the following conclusions:
\begin{itemize}
	\item The metallicity gradients found are negative ($\overline{\Delta [Z/H]}/\Delta \log(r/R_e) = -0.25 \pm 0.03$ for central galaxies and $\overline{\Delta [Z/H]}/\Delta \log(r/R_e) = -0.30 \pm 0.01$ for satellites; Fig. \ref{fig:met}). There is evidence of shallower (less negative) metallicity gradients with increasing stellar mass.
	
	\item Age gradients (Fig. \ref{fig:age}) are slightly positive ($\overline{\Delta \textrm{log (Age/Gyr)}/\Delta \log(r/R_e)}$ = 0.13 $\pm$ 0.03 for centrals and $\overline{\Delta \textrm{log (Age/Gyr)}/\Delta \log(r/R_e)}$ = 0.17 $\pm$ 0.01 for satellites). 
	
	\item The [$\alpha$/Fe] gradients are flat or slightly positive ($\overline{\Delta \alpha/Fe/\Delta \log(r/R_e)} = 0.01 \pm 0.03$ for centrals and $\overline{\Delta \alpha/Fe/\Delta \log(r/R_e)} = 0.08 \pm 0.01$ for satellites). We find [$\alpha$/Fe] gradients to have a weak negative correlation with stellar mass, showing more negative gradients with increasing mass (Fig. \ref{fig:alpha}). 
	
	\item The mean values of the gradients are consistent with a two-phase formation process. The metallicity gradients are shallower than the gradients expected from a pure dissipative collapse model (Table \ref{tab:mean_values}). The age and [$\alpha$/Fe] gradients are consistent with those expected for galaxies where the contribution from subsequent accretion events is significant.
	
	\item The stellar population gradients found for central galaxies are consistent with those found for satellite galaxies.
	
	\item The stellar population gradients of central and satellite galaxies show no significant difference as a function of halo mass. We find metallicity gradients of galaxies in the low-mass halo bins to have a potential positive correlation with stellar mass, showing less negative gradients with increasing mass. 

	\item The lack of significant difference with environment suggests that the stellar population gradients in the inner regions (r $<$ 2$R_e$) of passive galaxies have no significant dependence on their environment (Tab. \ref{tab:mean_values}). This result points to a similar formation and evolutionary history for the inner regions of central and satellite galaxies.
\end{itemize} 

In order to further test the formation process of central galaxies and reach a clearer understanding of how these galaxies form and the main processes that influence their evolution, observations of their stellar population gradients at larger radii are needed. 

\section*{Acknowledgements}
We thank the anonymous referee for their comments that
helped to improve this manuscript.\\
GS thanks SB and MM for their infinite patience and support.

The SAMI Galaxy Survey is based on observations made at the Anglo-Australian Telescope. The Sydney-AAO Multi-object Integral field spectrograph (SAMI) was developed jointly by the University of Sydney and the Australian Astronomical Observatory. The SAMI input catalogue is based on data taken from the Sloan Digital Sky Survey, the GAMA Survey and the VST ATLAS Survey. The SAMI Galaxy Survey is supported by the Australian Research Council Centre of Excellence for All Sky Astrophysics in 3 Dimensions (ASTRO 3D), through project number CE170100013, the Australian Research Council Centre of Excellence for All-sky Astrophysics (CAASTRO), through project number CE110001020, and other participating institutions. The SAMI Galaxy Survey website is \href{http://sami-survey.org/}{http://sami-survey.org/}.\\
SB acknowledges funding support from the Australian Research Council
through a Future Fellowship (FT140101166).
NS acknowledges support of an Australian Research Council Discovery Early Career Research Award (project number DE190100375) funded by the Australian Government and a University of Sydney Postdoctoral Research Fellowship.
MSO acknowledges the funding support from the Australian Research Council through a Future Fellowship (FT140100255).
JvdS is funded under Bland-Hawthorn's ARC Laureate Fellowship (FL140100278).
JJB acknowledges support of an Australian Research Council Future Fellowship (FT180100231).
IF gratefully acknowledges support from the AAO through
their distinguished visitor programme, as well as funding
from the Royal Society.
JBH is supported by an ARC Laureate Fellowship and an ARC Federation Fellowship that funded the SAMI prototype.

\bibliography{bibl.bib}
\appendix 
\section{Radial profiles of high-mass central galaxies.}
\begin{figure}[h]
	\centering
	\includegraphics[height=9cm, width=9cm, keepaspectratio]{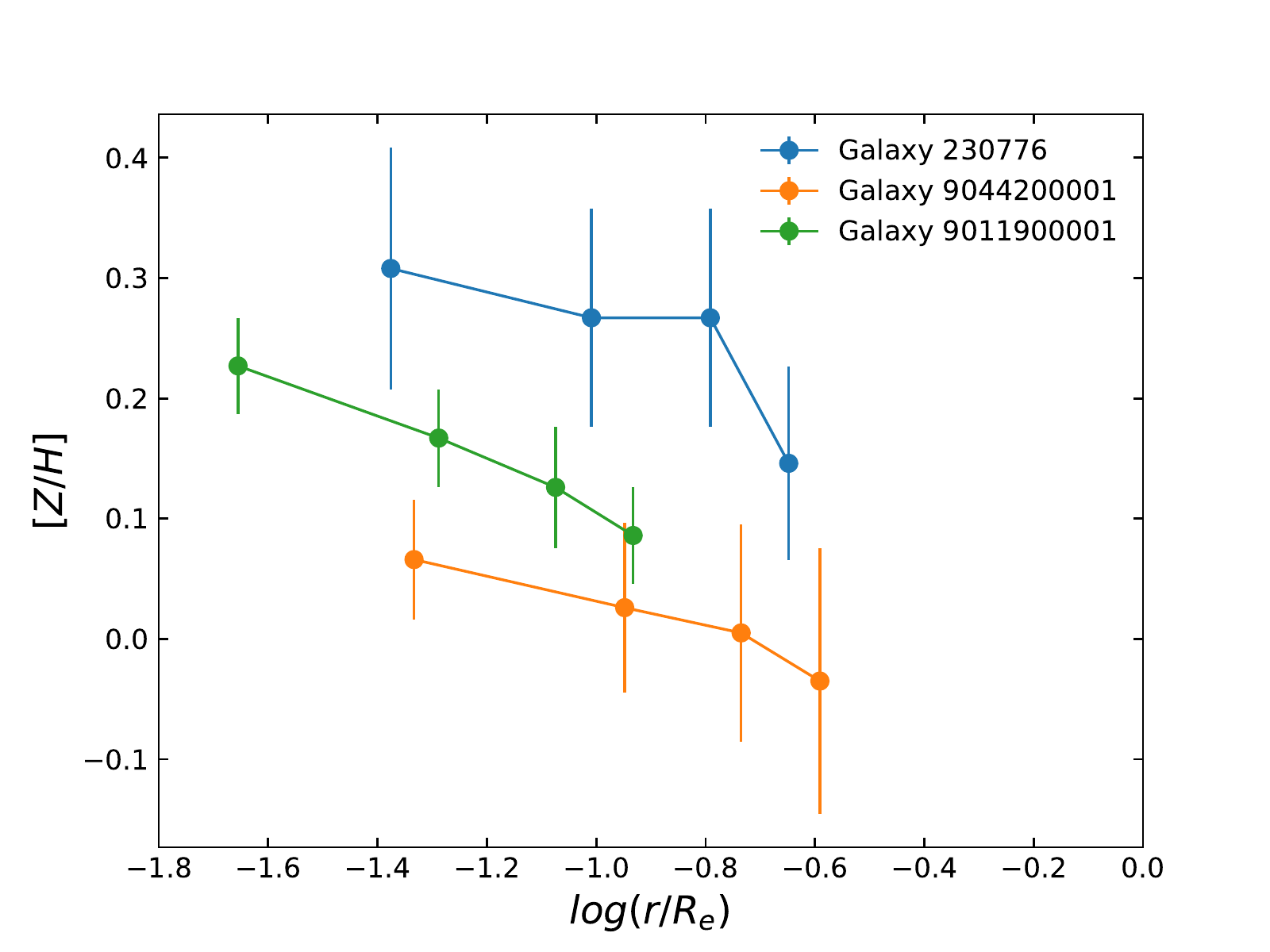}  
	\caption{Radial metallicity profiles of the three high-mass central galaxies with measurements only within 0.3 $R_e$. We retain these galaxies in the final sample (we show their radial metallicity profiles for transparency).  }
	\label{fig:cen_radial_prof}
\end{figure}

Three high-mass ($\log_{10} (M_*/ M_{\odot}) > 11.4$) central galaxies (2 in clusters and one in the GAMA groups, are very large ($R_e > 25^{\prime\prime}$) compared to the $15^{\prime\prime}$ diameter SAMI field of view. Though their measurements only cover a small portion of the galaxy, up to 0.26 Re, they still meet our selection criteria. We illustrate their radial metallicity profiles in Fig. \ref{fig:cen_radial_prof}.\\

\section{Stellar population maps}
\begin{figure}[!ht]
	\centering
	\includegraphics[trim=0.5cm 1.5cm 0.5cm 1.8cm,clip=true,width=4.78cm]{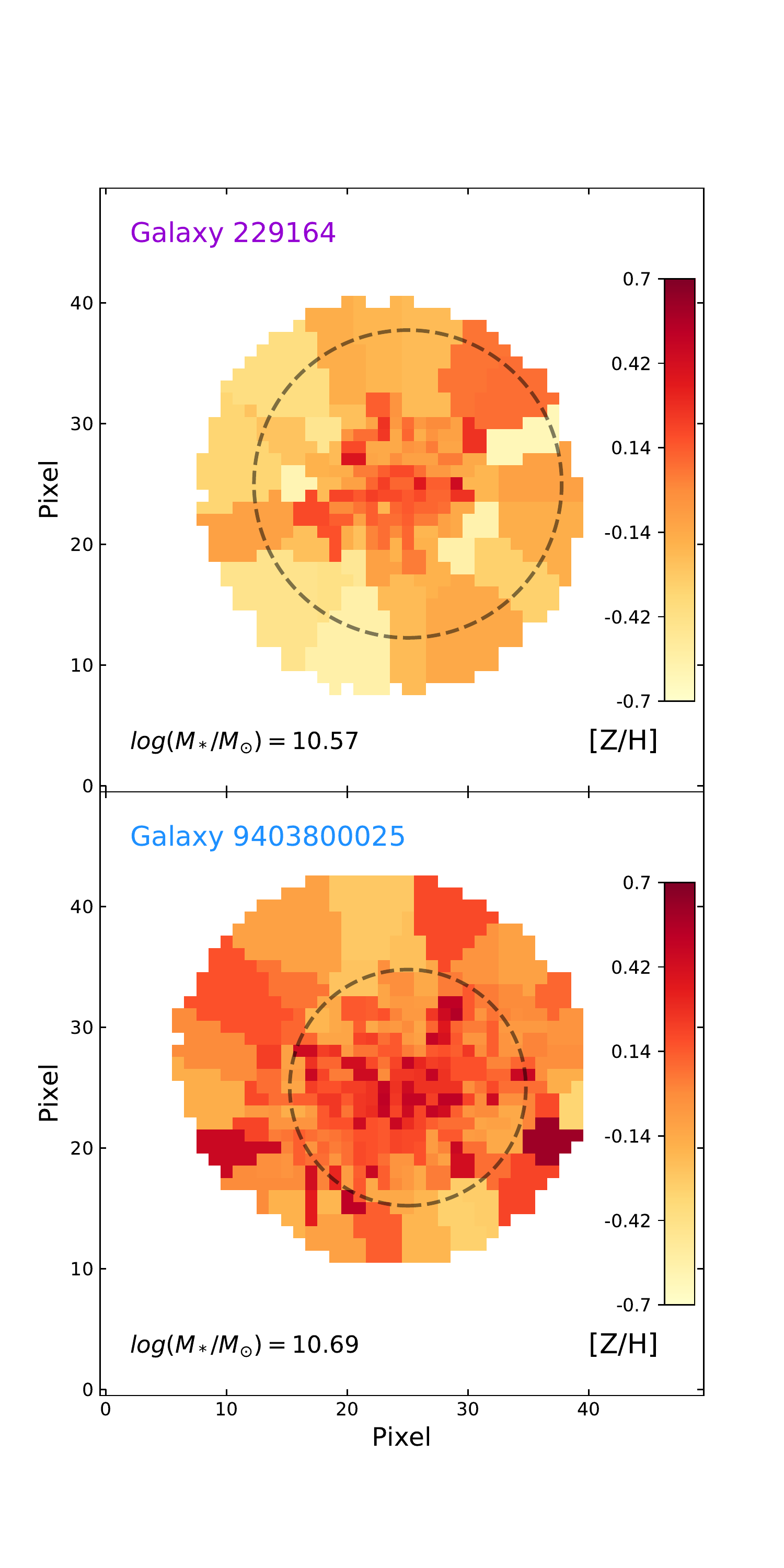}\includegraphics[trim=0.5cm 1.2cm 0.5cm 2.5cm,clip=true,width=7.5cm]{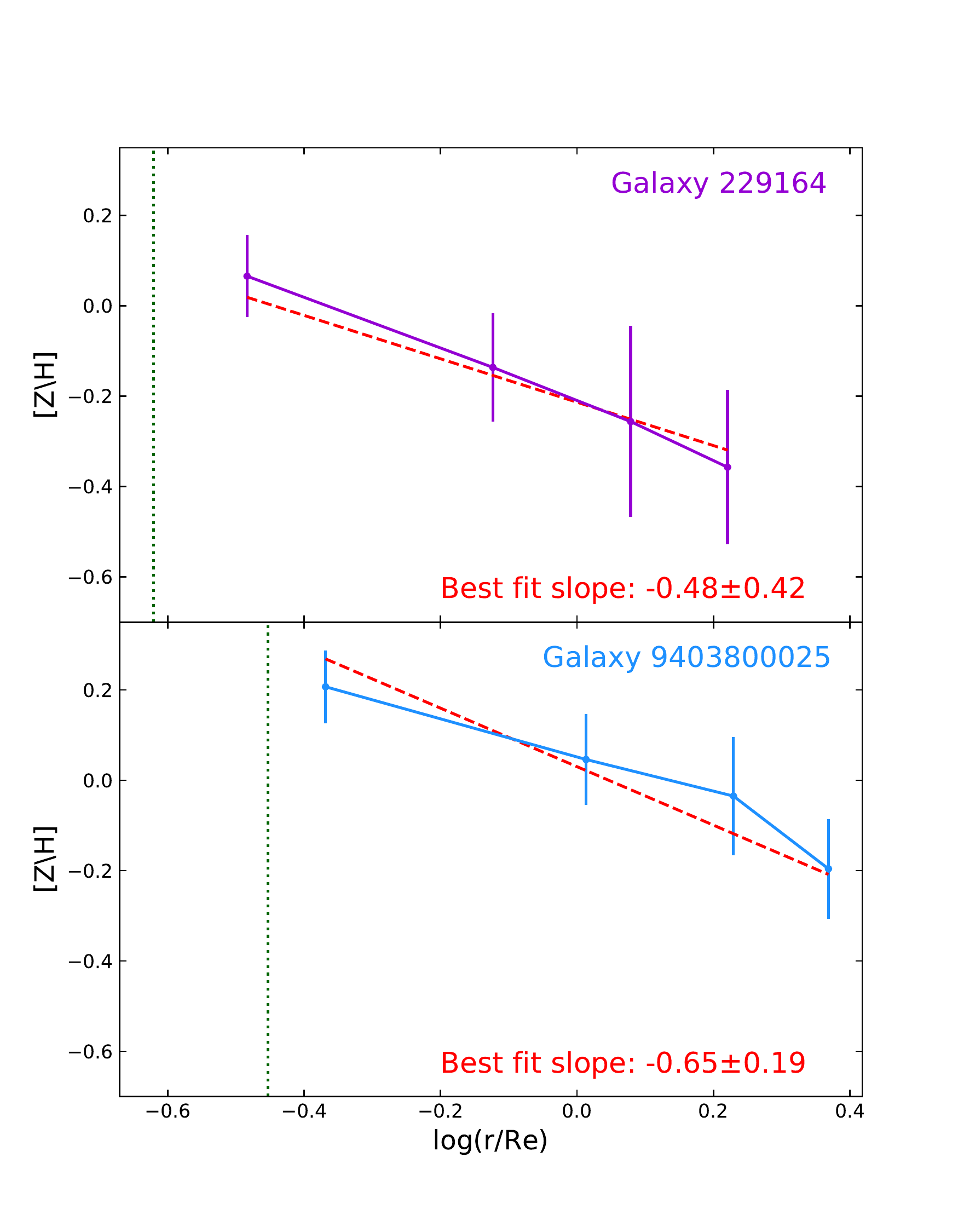}
	\caption{Example metallicity maps and radial metallicity profiles for 2 typical galaxies: central galaxy and satellite galaxy. Radial profiles are color-coded as follows: purple for central galaxies and blue for satellite galaxies. Stellar mass is given in the bottom right of the first panel. $R_e$ is shown in the first panel as a dashed black circle. PSF extent is shown in the radial profile as the green dotted line. The stellar metallicity decreases going toward the outskirts of the galaxy.}
	\label{fig:met_example}
\end{figure}

\begin{figure}[!ht]
	\centering
	\includegraphics[trim=0.5cm 1.5cm 0.5cm 1.8cm,clip=true,width=4.78cm]{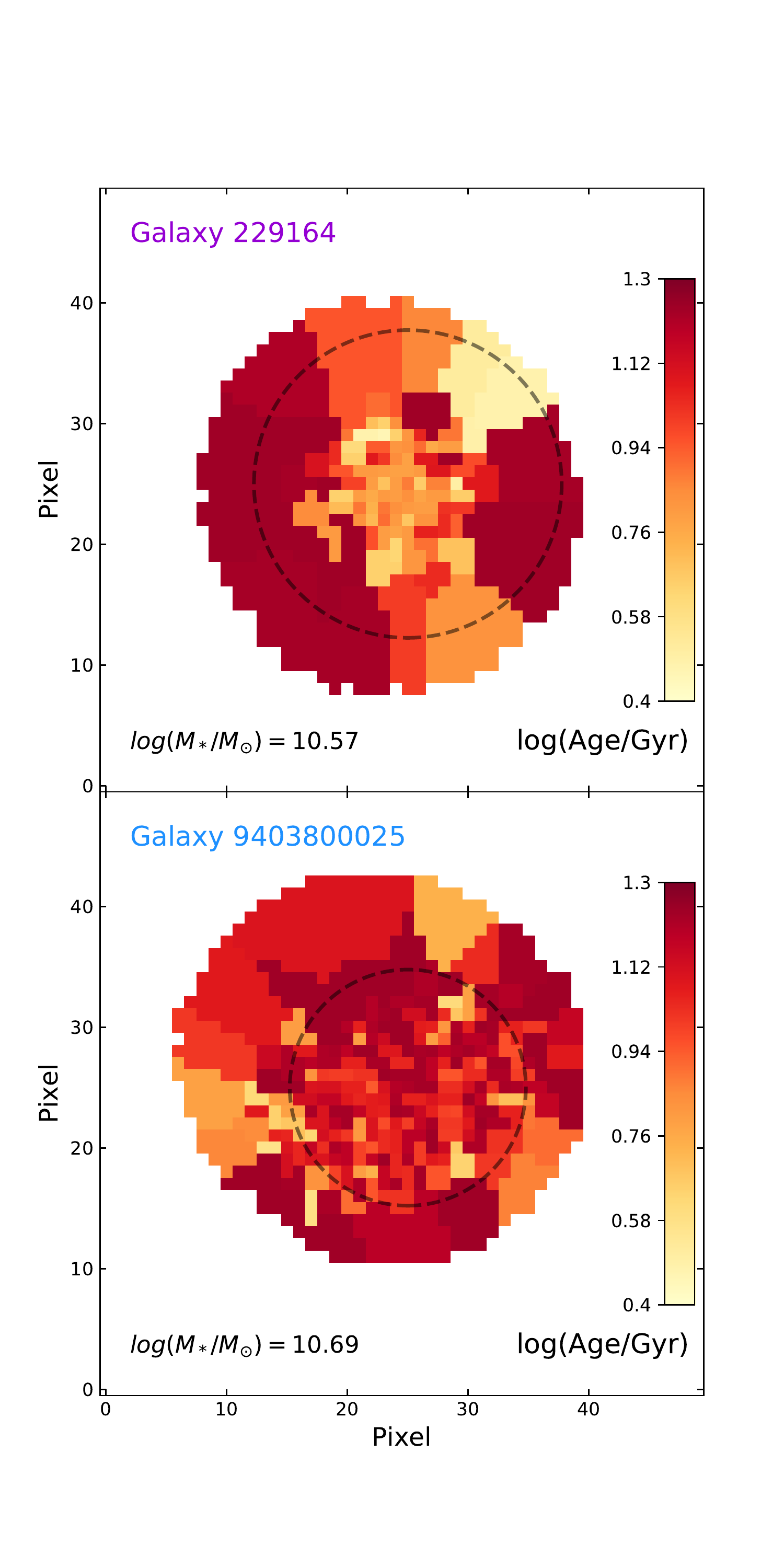}\includegraphics[trim=0.5cm 1.2cm 0.5cm 2.5cm,clip=true,width=7.5cm]{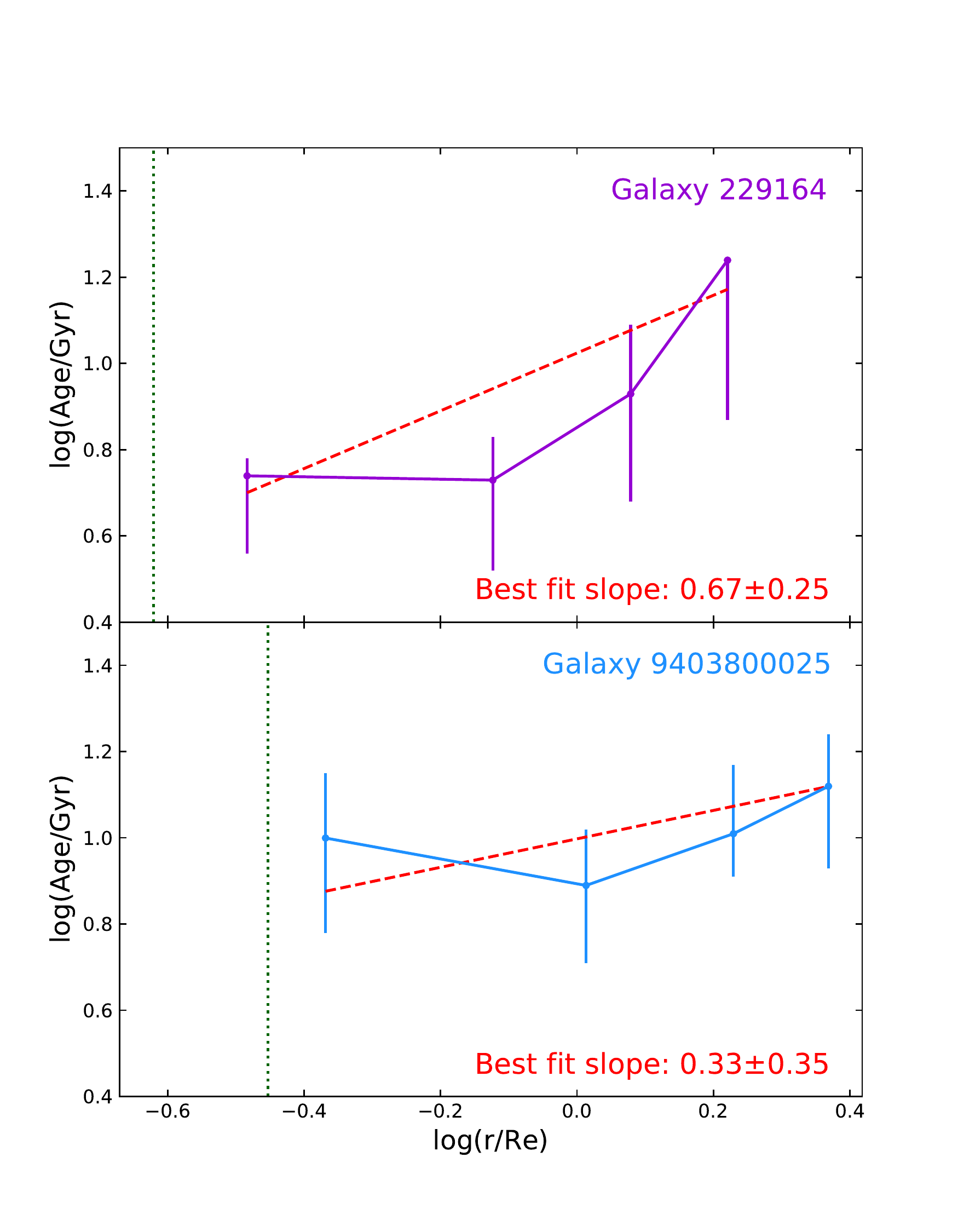} 
	\caption{Example age maps and radial age profiles for 2 typical galaxies: central galaxy and satellite galaxy. Radial profiles are color-coded as follows: purple for central galaxies and blue for satellite galaxies. Stellar mass is given in the bottom right of the first panel. $R_e$ is shown in the first panel as a dashed black circle. PSF extent is shown in the radial profile as the green dotted line.}
	\label{fig:age_example}
\end{figure}

\begin{figure}[!ht]
	\centering
	\includegraphics[trim=0.5cm 1.5cm 0.5cm 1.8cm,clip=true,width=4.78cm]{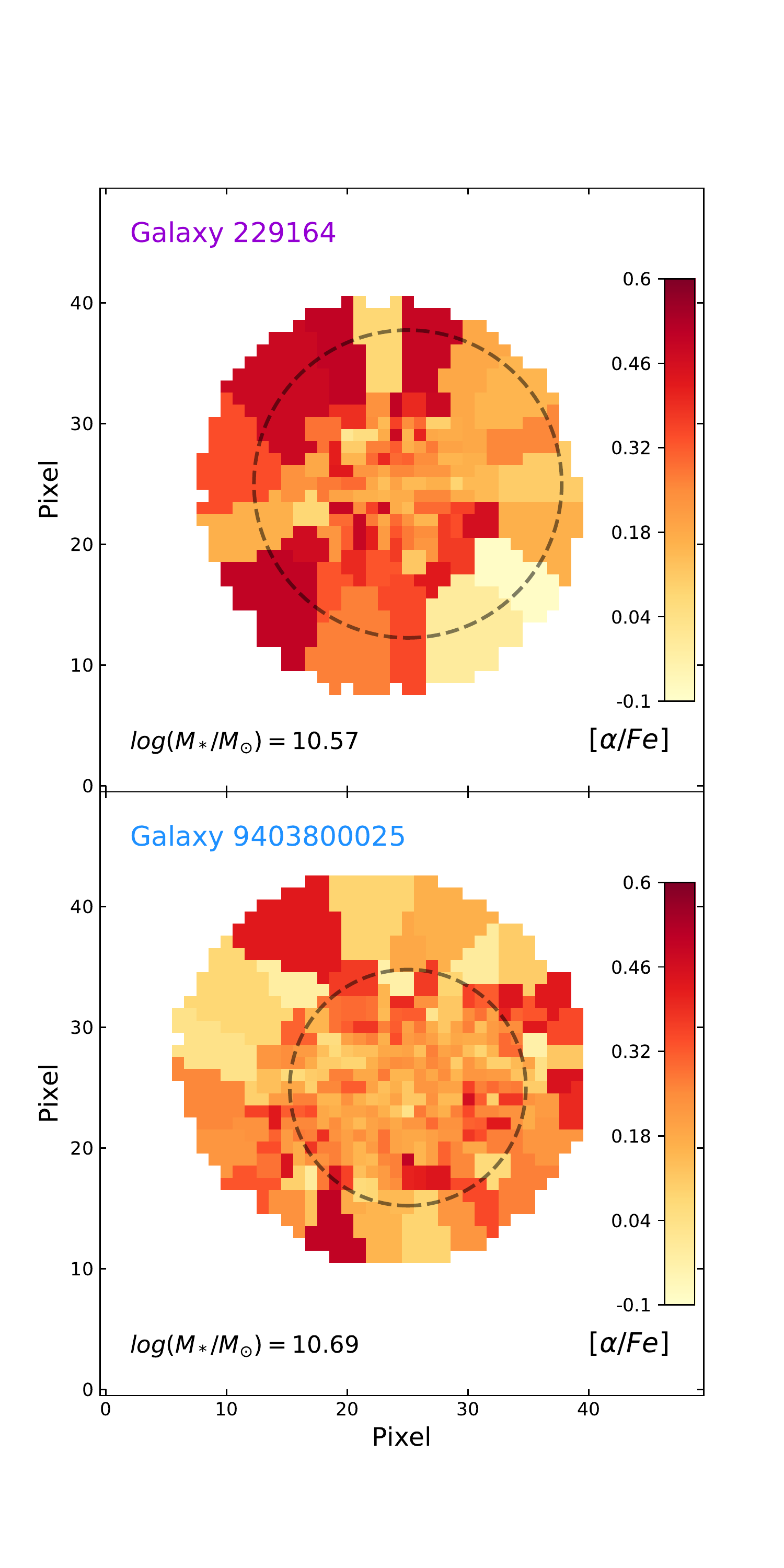}\includegraphics[trim=0.5cm 1.2cm 0.5cm 2.5cm,clip=true,width=7.5cm]{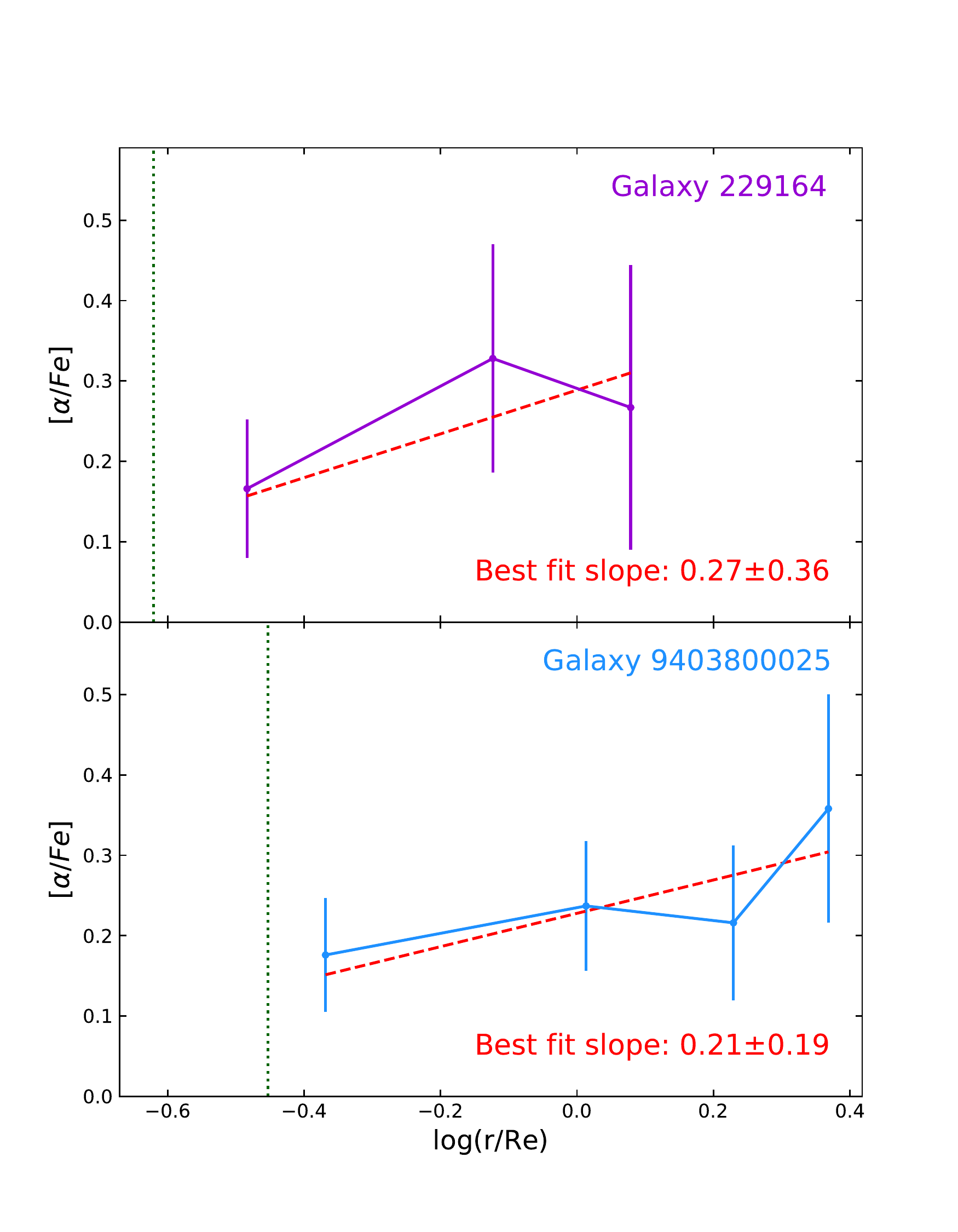} 
	\caption{Example [$\alpha$/Fe] maps and radial profiles for 2 typical galaxies: central galaxy and satellite galaxy. Radial profiles are color-coded as follows: purple for central galaxies and blue for satellite galaxies. Stellar mass is given in the bottom right of the first panel. $R_e$ is shown in the first panel as a dashed black circle. PSF extent is shown in the radial profile as the green dotted line.}
	\label{fig:alpha_example}
\end{figure}

Example stellar population (metallicity, age and [$\alpha$/Fe]) binned maps and radial profiles are shown in Figs. \ref{fig:met_example} -\ref{fig:alpha_example} for 2 galaxies (1 central and 1 satellite galaxy). Stellar population binned maps have been created specifically for these 2 examples and have not been used in this study.
\\
\section{AGN contribution}
\begin{figure}[!ht]
	\centering
	\includegraphics[trim=0cm 6cm 0.5cm 6cm,clip=true, width=8.5cm, keepaspectratio]{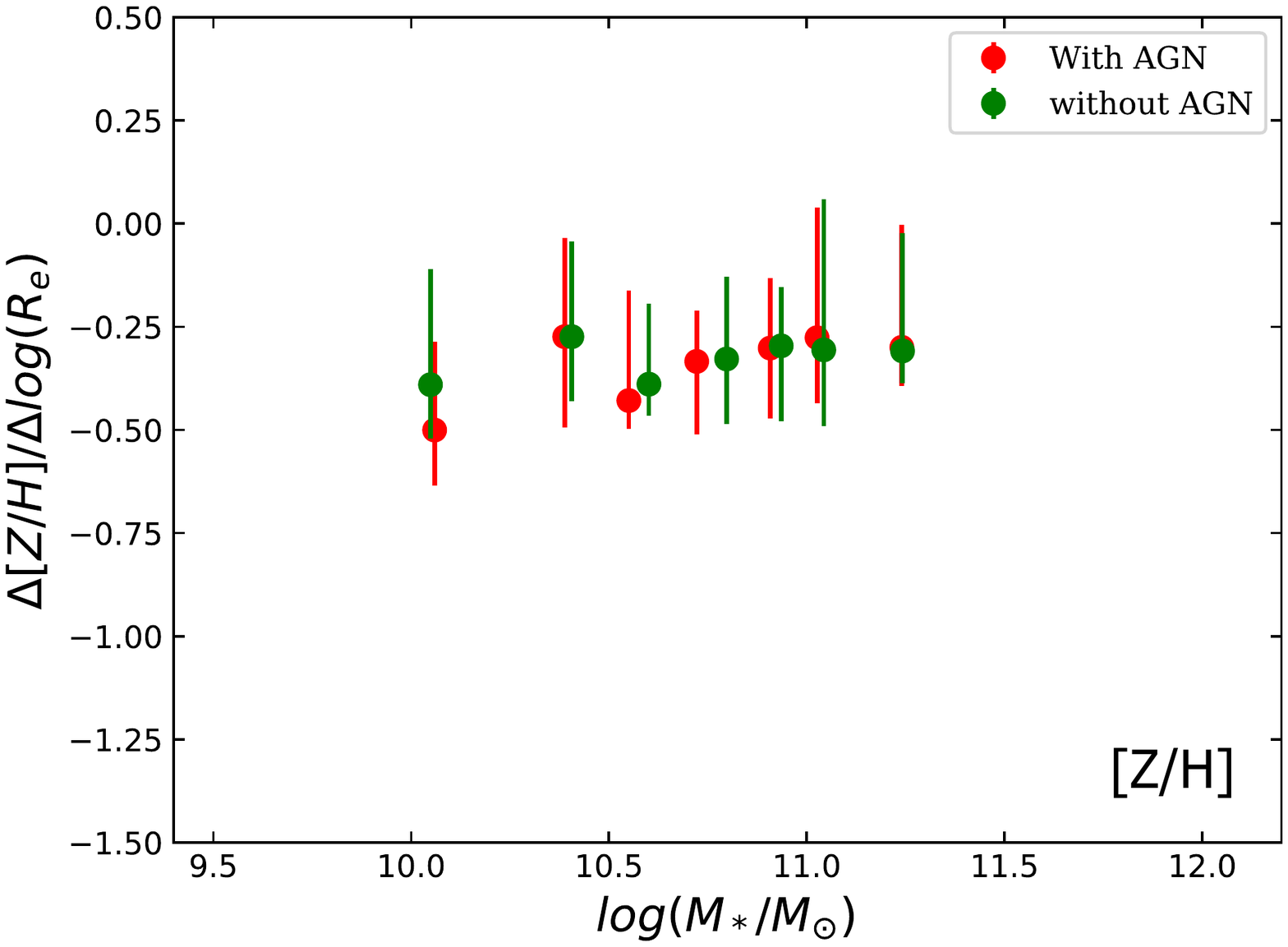}  \includegraphics[trim=0cm 6cm 0.5cm 6cm,clip=true, width=8.5cm, keepaspectratio]{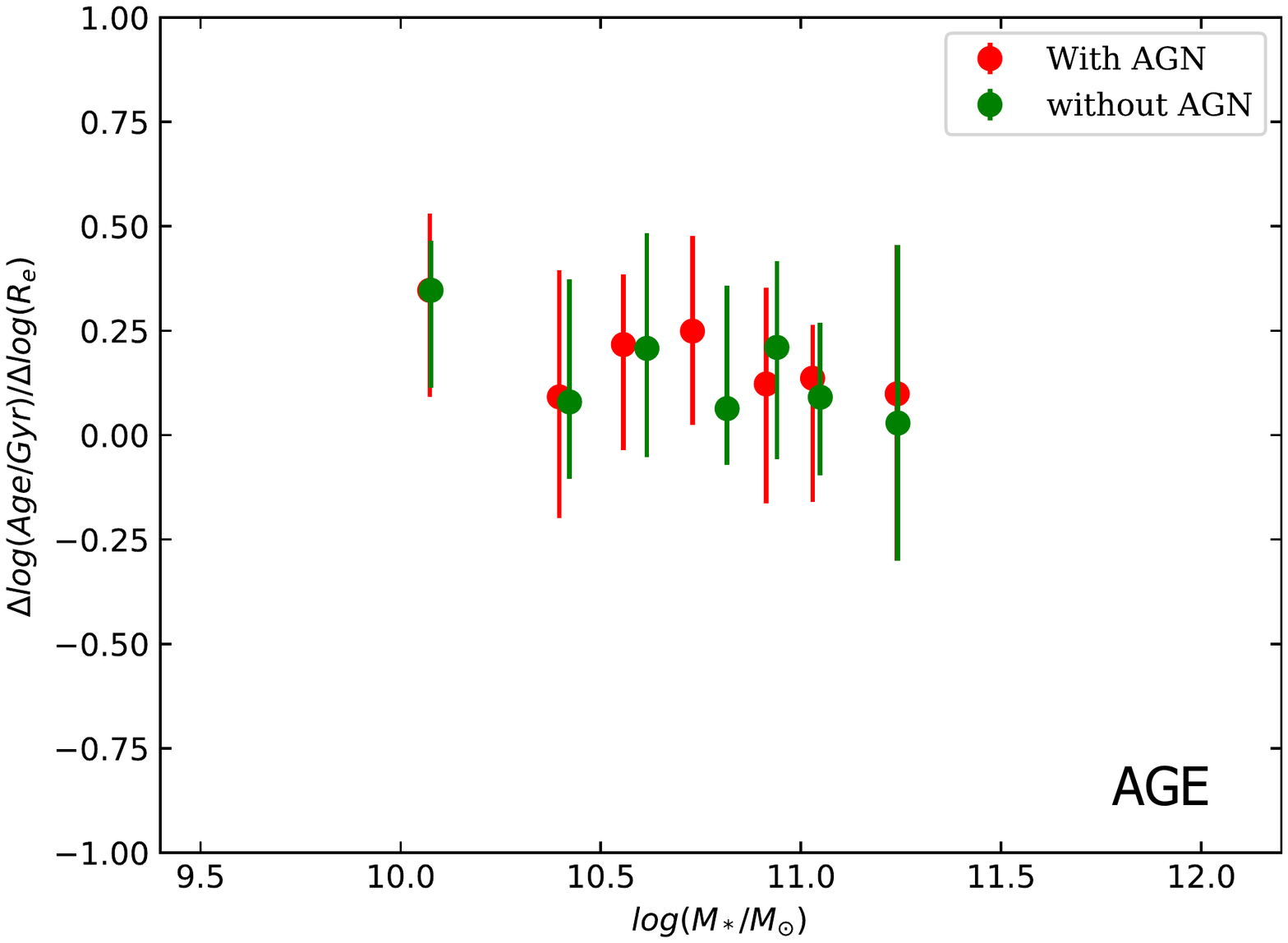}
	\caption{Left plot: metallicity gradients for all of the GAMA galaxies in our sample and the metallicity gradients derived excluding galaxies with AGN emission, binned by mass, as a function of $\log_{10} (M_*/ M_{\odot})$. Errorbars show the $25^{th}$ and $75^{th}$ percentile in each bin. Right plot: age gradients as per left plot. There is no significant difference between the gradients in the two samples.}
	\label{fig:AGN_z}
\end{figure}

Since AGN have strong, very bright emission lines in their centers that may bias stellar 
population fitting, we investigate the importance of this contribution by comparing gradients 
of a sample of galaxies that includes galaxies known to have AGN activity, in the GAMA regions, 
with non-AGN GAMA galaxies (168 galaxies with AGN out of 332 in our sample; \citealt{Schaefer2017}).
As shown in Fig. \ref{fig:AGN_z}, there is no significant difference in the metallicity and age gradients between the two samples. We therefore conclude that there is no significant influence from AGN emission on our stellar population measurements and include all galaxies in our sample.
\end{document}